\documentclass[prd,twocolumn,amsmath,nobibnotes,nofootinbib,superscriptaddress,preprintnumbers]{revtex4}

\usepackage{bm} \usepackage{graphicx}

\usepackage{subfigure}
\newcommand{\be}{\begin{equation}}
\newcommand{\bea}{\begin{eqnarray}}
\newcommand{\ee}{\end{equation}}
\newcommand{\eea}{\end{eqnarray}}

\newcommand{\cG}{\mathcal G}
\newcommand{\cV}{\mathcal V}

\newcommand{\cR}{\mathcal R}

\newcommand{\cN}{\mathcal N}

\begin{document}
 
\title{Field dynamics and tunneling in a flux landscape}

\author{Matthew C. Johnson}
\email{mjohnson@theory.caltech.edu}
\affiliation{California Institute of Technology, Pasadena, CA 91125, USA} 
\author{Magdalena Larfors}
\email{magdalena.larfors@fysast.uu.se}
\affiliation{Institutionen f\"or Fysik och Astronomi, Box 803, SE-751 08 Uppsala, Sweden. }

\begin{abstract}
We investigate field dynamics and tunneling between metastable minima in a landscape of Type IIB flux compactifications, utilizing monodromies of the complex structure moduli space to continuously connect flux vacua. After describing the generic features of a flux-induced potential for the complex structure and Type IIB axio-dilaton, we specialize to the Mirror Quintic Calabi--Yau to obtain an example landscape. Studying the cosmological dynamics of the complex structure moduli, we find that the potential generically does not support slow-roll inflation and that in general the landscape separates neatly into basins of attraction of the various minima. We then discuss tunneling, with the inclusion of gravitational effects, in many-dimensional field spaces. A set of constraints on the form of the Euclidean paths through field space are presented, and then applied to construct approximate instantons mediating the transition between de Sitter vacua in the flux landscape. We find 
 that these instantons are generically thick-wall and that the tunneling rate is suppressed in the large-volume limit. We also consider examples where supersymmetry is not broken by fluxes, in which case near-BPS thin-wall bubbles can be constructed. We calculate the bubble wall tension, finding that it scales like a D- or NS-brane bubble, and comment on the implications of this correspondence. Finally, we present a brief discussion of eternal inflation in the flux-landscape.
\end{abstract}

\preprint{UUITP-09/08, CALT-68.2687}

\date{\today}

\maketitle

\section{Introduction}

The string theory landscape is a large collection of four-dimensional, low-energy effective field theories that are obtained by compactifying string or M-theory on an internal manifold~\footnote{For an introduction to the string landscape, see~\cite{Susskind:2003kw}.}. These theories arise as vacua in a complicated potential that typically depends on hundreds of parameters. Because there are usually many ways to stabilize any given internal manifold (and many choices of the internal manifold itself), the number of vacua can be extremely large. 

It is natural to start an investigation of the landscape by looking at the properties of vacua. What kind of four-dimensional theories occur in the landscape? Is any kind of effective space-time preferred? What is the distribution of the effective cosmological constant and the scale of supersymmetry breaking? Given the great number of expected vacua, a statistical treatment of these questions is necessary. Work in this direction has been carried out by e.g.~\cite{Douglas:2003um,Denef:2004cf,Denef:2004ze}.

However, to gain a full understanding of the history and large-scale structure of the universe, studying the vacua of the potential is not enough. In order to describe the cosmological history of our universe, including an epoch of slow-roll inflation, one must at least have information about the potential in the vicinity of a viable minimum. Furthermore, many vacua in the landscape (including all de Sitter vacua) are not global minima. Thus they are only metastable when quantum mechanical corrections are taken into account, leading to a slow migration between vacua. From the four-dimensional perspective, this will proceed through the nucleation of a bubble of a lower energy phase ( the `true' vacuum) inside a region of a higher energy phase (the `false' vacuum), a process which has been discussed extensively in the literature, starting with the works of Coleman and collaborators~\cite{Callan:1977pt,Coleman:1977py,Coleman:1980aw}.

Suppose that at some stage in the history of our Universe, a local
region settled into a de Sitter vacuum in the landscape. If the probability of a bubble forming in each Horizon volume is low (more precisely, the decay probability per unit four-volume must be less than $H_F^4$), it follows that some volume of the original phase will always survive. This phenomenon is known as eternal inflation, and together with the landscape, it implies that not only are there a vast number of low-energy limits of string theory, but that many (or perhaps all)  {\em are actually realized} in different spatiotemporal regions. In light of this, the predictions made from the landscape for low-energy physics can only be statistical. How the statistical distributions of various physical properties can be mapped out is at this point not completely understood (see~\cite{Winitzki:2006rn,Guth:2007ng}). However, to carry out this program, it will be necessary to study the allowed types and properties of transitions between vacua~\cite{Aguirre:2006na}. 

The transition probability is determined by the topographic features of the landscape, e.g. by the height and width of the potential barriers. These features of the landscape are in general difficult to study, since they are highly model dependent. We need an explicit expression for the potential to determine how (and if) vacua are continuously connected by potential barriers. However, classes of compactifications can have common topographic features, and here we will focus on one such example, namely the topography of flux compactifications. 

One way of obtaining stable string theory compactifications on internal manifolds is to let generalized, higher-dimensional magnetic fluxes pierce non-trivial cycles of the internal manifold. Such fluxes are present as $p-$form fields in the string theory that is compactified. Given the flux, energy is required to change the size of the pierced cycle. Hence, the fluxes introduce a potential for the parameters, or moduli, that determines the shape of the compactification manifold. The moduli can be stabilized at a minimum of the potential, perhaps yielding a  phenomenologically acceptable four-dimensional theory. There are several reviews on flux compactifications, e.g.~\cite{Grana:2005jc,Douglas:2006es}.

Generally, there are symmetries of the compactification manifold, known as monodromy transformations, that act on the non-trivial internal cycles (for examples see~\cite{Candelas:1990pi,Candelas:1990rm} or~\cite{Hori:2003ic}). Adding flux, as we describe in more detail later, these transformations will not be symmetries of the resulting potential, but can instead be viewed as a way to continuously construct the potential between minima with different flux configurations. The topography of such continuously connected series of vacua can be investigated, and it is reasonable to expect that these features are somewhat universal among different flux compactifications. In this paper, we focus on series of minima in the flux potential for complex structure moduli of compactifications of Type IIB string theory. As an example, we will study the flux minima for the complex structure modulus of the Mirror Quintic, where it has been shown that series of continuously connected minima exi
 st~\cite{Danielsson:2006xw}. 

In doing so, we will be able to explicitly describe the transitions between flux vacua in terms of a four-dimensional low-energy theory. This is an important property to establish, and has been a topic of some controversy in the past~\cite{deAlwis:2006cb,deAlwis:2006am}. In addition, our picture of transitions in the flux landscape illustrates a number of features not present in the flux lattice of the Bousso--Polchinski landscape~\cite{Bousso:2000xa}, and we are able to clarify the connection between near-BPS domain walls~\cite{Ceresole:2006iq} and D- and NS-brane bubbles. These results  complement the recent analysis of Ref.~\cite{Dine:2007er}. 

The paper is organized as follows. Sec.~\ref{sec:landscape} introduces the string landscape and explains how cycle monodromies affect the topography of the landscape. The following section is an in depth investigation of the properties of the Mirror Quintic example landscape that will be used throughout the paper. The classical field dynamics of the flux sector is discussed. We then focus on tunneling, and the stability of flux vacua. Sec.~\ref{sec:tunneling} contains a general discussion of tunneling in multi-dimensional field spaces. In Sec.~\ref{sec:flux_tun} we calculate tunneling amplitudes in the Mirror Quintic, and discuss the interpretation of the four-dimensional bubbles. We then discuss eternal inflation in Sec.~\ref{sec:et_inf}. Finally, our results are summarized and discussed in Sec.~\ref{sec:conclusions}. Our notation and some technical computations are found in Appendix~\ref{ApB}.

\section{Basics of the string theory landscape}
\label{sec:landscape}

\subsection{Scales in string theory}
\label{sec:scales}
Most of the phenomenologically interesting computations in string theory lie within the realm of a low-energy effective theory, ten-dimensional supergravity. There are two approximations that must be made in order to use this framework. First, we must restrict ourselves to energy scales much lower than $M_s$, the energy splitting between string states in ten-dimensional Minkowski space. This allows us to neglect the internal degrees of freedom of strings. We must also work at weak string coupling, $g_s$, which suppresses string loop contributions to the effective action. The string coupling is a dynamical field, related to the dilaton field by $g_s = e^{\phi}$, and so this is a restriction on its range of variation. In addition to the graviton, the theory  generally includes the dilaton, a number of $p$-form fields, and extended objects known as D$p$-branes. The exact field content depends on the string theory whose low-energy limit we are interested in.

In order to make contact with four-dimensional physics, we must compactify six of the original ten dimensions. This dimensional reduction requires us to specify the manifold that the six extra dimensions are compactified to. In the absence of flux and branes, this manifold can be characterized by its topology and a set of metric deformations known as moduli fields, which at this level are exactly massless (i.e. it incurs no energy to change their values). In this paper, we  compactify on Calabi--Yau manifolds, which allow us to preserve some supersymmetry in four dimensions. The metric deformations then fall into two different cohomology classes: K{\"a}hler and complex structure. Schematically, the K{\"a}hler moduli correspond to a scaling of the volume and the complex structure moduli correspond to a change in the shape of the manifold.

Reducing the volume of the six dimensional manifold, we obtain an effective four dimensional theory that contains the four-dimensional graviton and an infinite set of fields for each of the moduli of the underlying compactification (known as the Kaluza--Klein, or KK,  tower). Each set of fields contains a  massless mode as well as a sequence of modes whose masses are determined by the eigenvalues of the six-dimensional Laplacian, and therefore by the volume of the internal manifold. This introduces a mass scale into the four-dimensional theory, $M_{KK}$, below which we can keep only the massless mode.

We now collect a number of results from Appendix~\ref{ApB} for the reader's convenience. We define the internal volume as $\cV = \rho_I^{3/2} \tilde{\cV} $, where $\tilde{\cV} \equiv (2 \pi)^6 \alpha'^3$ and $\rho_I$ is a dimensionless (K{\"a}hler) modulus determining the overall scale of the internal manifold. We then have the following ratios of scales
\begin{equation}\label{eq:scales}
\frac{M_s}{M_p} = g_s^{1/4} \rho_I^{-3/4}, \ \ \frac{M_{KK}}{M_p} = \rho_I^{-1}, \ \ \frac{M_{KK}}{M_s} = (g_s \rho_I)^{-1/4}.
\end{equation}
For weak string coupling ($g_s < 1$), and large volume ($\rho_I \gg 1$), we will have a string scale that is much lower than the four-dimensional Planck mass $M_p$~\footnote{The hierarchy can be larger for warped internal manifolds \cite{DeWolfe:2002nn}. It is important that all scales are measured in the same frame, e.g. the ten-dimensional string frame, but the ratios are of course frame independent.}. Depending on the relative size of $g_s$ and $\rho_I$, we could find that the KK scale is either above or below the string scale.

\subsection{Moduli stabilization and the landscape}\label{modstabland}

The massless moduli introduced above are not phenomenologically acceptable; they introduce long-ranged so-called fifth forces and cause problems with cosmology. It is therefore necessary for these fields to be stabilized at a relatively high mass scale in a realistic low-energy theory. In order to fix the moduli, one must consider adding fluxes, branes, $\alpha'$ corrections, and/or non-perturbative corrections. Because of these complications, successful stabilization of all the moduli in a string theory compactification has been achieved only relatively recently. Examples of stabilized theories in the context of Type IIB and Type IIA string theories are found in~\cite{Kachru:2003aw,Balasubramanian:2005zx,Giddings:2001yu,Conlon:2005ki} and~\cite{DeWolfe:2005uu,Villadoro:2005cu} respectively. The interesting conclusion reached in this analysis is that not only are there many different manifolds to compactify on, but many vacua exist for the moduli of a given compactification. 
 This potentially large collection of vacua has been dubbed the string theory landscape~\cite{Susskind:2003kw}, and the existence of many different consistent four-dimensional low-energy limits of the original ten-dimensional supergravity and a specified compactification has profound implications for making observational connections with UV physics.

Further complicating matters, the moduli are in general only fixed at metastable vacua of the landscape. The potential barriers separating vacua can be penetrated by quantum mechanical tunneling of the moduli fields. The stability of a given vacuum is determined by the topography of the landscape surrounding it: the distance to the next minimum, barrier heights, etc. Therefore, to understand the dynamics on the landscape we must investigate its topography. In the following we will focus on the topography of the sector of the landscape that is fixed by fluxes.

\begin{figure*}[tb]
\begin{center}
\includegraphics[height=5.8cm]{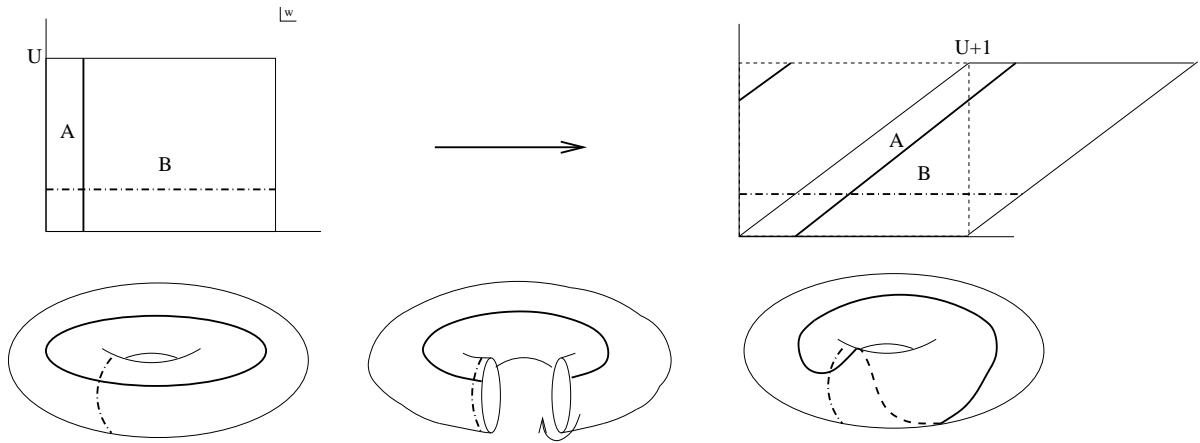}
\end{center}
\caption{Here, we depict one of the monodromies of a two-dimensional torus. On the top left, the torus can be represented by a region of the complex plane with opposite edges of the rectangle identified. Twisting the torus by an angle of $2 \pi$ (bottom) is equivalent to taking $U \rightarrow U + 1$ (top). The result is a torus that is identical in shape, but with transformed cycles (bottom right).}
\label{fig:torus}
\end{figure*}

We can gain much intuition about the topography of this landscape sector by looking at the example of a two dimensional torus (to learn more about the torus, see~\cite{Nakahara:2003nw}). As shown in Fig.~\ref{fig:torus}, a torus can be represented as a lattice on the complex plane with the identifications
\begin{equation}\label{eq-zidentify}
w \rightarrow w + m + n U,
\end{equation}
where $m$ and $n$ are integers and the complex parameter $U \equiv A /B$ is the ratio of periods of the non-contractible one-cycles $a$ and $b$. The coordinate $w$ is dimensionless, with the overall scale measured by the volume of the torus. There are then two moduli, the complex structure modulus $U$, and the overall volume (just $AB$ for a rectangular torus) which is a K{\"a}hler modulus that we neglect for now.

When $U$ has a non-zero real part, it is no longer rectangular in the original lattice. In the three-dimsional analogue, the torus is "twisted" by the angle $\theta \equiv 2 \pi Re(U)$ (see Fig.~\ref{fig:torus}). Not all twists generate new tori. Twisting by an angle of $2 \pi$, and applying the identifications Eq.~\ref{eq-zidentify}, we see that the new torus cover the same lattice points as the original one. Thus the modular transform $U \rightarrow U + 1$ is a symmetry on $U$ space. This transformation, together with $U \rightarrow -1/U$, generate the full symmetry group of the torus, PSL(2,Z). The complex structure moduli space of the torus consists of all points in $U$ space not related by these two identifications.

Note that the modular transforms affect the non-contractible cycles of the torus. Under $U \rightarrow U + 1$ the $b$-cycle is unchanged, but the $a$-cycle undergoes the  {\it monodromy} transform $a \rightarrow a + b$ (see Fig.~\ref{fig:torus}). Similarly, under $U \rightarrow -1/U$, the two cycles are interchanged. Thus, the modular transforms on $U$ space induce two monodromy transformations on the cycles, or equivalently their periods 
\begin{eqnarray}
\left( \begin{array}{c}
A\\
B \end{array} \right) 
\rightarrow {\bf T}_{i} \left( \begin{array}{c}
A\\
B \end{array} \right) ,
\end{eqnarray}
as described by the monodromy matrices 
\begin{eqnarray}
{\bf T}_{0} = \left( \begin{array}{cc}
1 & 1\\
0 & 1 \end{array} \right) , \ \ \  
{\bf T}_{1} = \left( \begin{array}{cc}
0 & -1 \\
1 & 0  \end{array} \right).
\end{eqnarray}

Now imagine that a constant magnetic field exists on the torus. Because of the non-trivial topology, the field can be present without sources, but the total flux through each of the cycles obeys a Dirac quantization condition. Confining flux lines to a region in space costs energy, meaning that a potential is induced for the size of the cycles. We therefore induce a potential on the torus' complex structure moduli space. Also, in the presence of flux, the twisting described above comes with a price in energy due to the fact that the lines of flux must twist as well in order to remain unbroken. 

The most important property to note about the potential in this example is that while monodromies preserve the geometrical properties of the manifold, they will not preserve the form of the potential, implying that it is multi-valued on the complex structure moduli space. We can twist as much as we like, producing a potential that resembles a spiral staircase (with the various sheets matched across branch cuts) as we wind the lines of flux more and more.

Twisting effectively adds components of flux to the cycle in the direction of the twist, and taking $U \rightarrow - 1/ U$ changes the cycle a flux is wrapping. This means that we can view monodromy transformations as a way to move continuously between different flux configurations on the torus. It is possible to connect many, but not all flux configurations in this way (e.g., starting with a flux configuration $(F,G)$ around the two basis cycles, we can never reach $2(F,G)$, since the matrix $2 I$ has determinant 2 and cannot be in SL(2,Z)).

It has been shown by one of the authors that there are continuous sequences of vacua related by monodromies in Type IIB flux compactifications~\cite{Danielsson:2006xw}, and possibly an infinite number of them~\cite{Chialva:2007sv}. The simple example of the torus captures many of the features found in such Type IIB flux landscapes, and should be kept in mind as we discuss more complicated internal manifolds. 

\section{Mirror Quintic Calabi--Yau}
\label{sec:MQ}

The addition of fluxes on the Mirror Quintic Calabi--Yau (see~\cite{Candelas:1990rm} for a detailed description of the properties of this manifold) generates an example landscape of continuously connected Type IIB flux vacua. This model was studied previously in~\cite{Danielsson:2006xw}, to which we refer the reader for notation and additional details. The Mirror Quintic possesses 101 K{\"a}hler moduli and a single complex structure modulus, and provides a simple model for studying the structure of the potential on complex structure moduli space. There are four basis 3-cycles that can be pierced by flux, inducing a potential for the complex structure modulus and the Type IIB axio-dilaton, and we will assume in what follows that non-perturbative effects or $\alpha'$ corrections fix the K{\"a}hler moduli. We expect this analysis of the complex structure moduli space to carry over to models with fixed K{\"a}hler moduli, and provide insights into the topographic features of the f
 ull landscape.

\subsection{The moduli space, flux, and monodromies}

\begin{figure}[tb]
\begin{center}
\includegraphics[height=4.8cm]{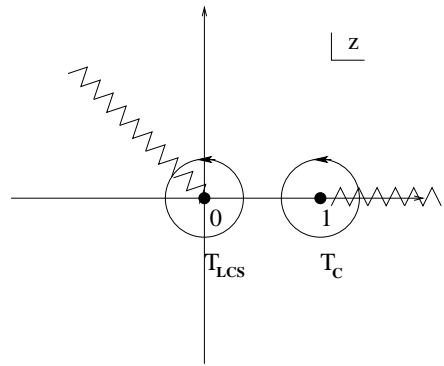}
\end{center}
\caption{ The complex structure moduli space of the Mirror Quintic. The plot shows two of the three singular points: the large complex structure point $z=0$ and the conifold point $z=1$. The two branch cuts correspond to the large complex structure and the conifold monodromies.}
\label{fig:MQ}
\end{figure}

The periods, $\Pi = (\Pi_1,\Pi_2,\Pi_3,\Pi_4)^T$, over the four basis three-cycles specify the geometry of the complex structure moduli space of the Mirror Quintic, as described in Appendix~\ref{ApB}. There is one complex structure modulus $z$, and the  complex structure moduli space is the complex plane shown in  Fig.~\ref{fig:MQ}. This space has three singularities, the large complex structure (LCS) point ($z=0$), the conifold point ($z=1$) and the Gepner point ($z=\infty$). There is a branch cut emanating from each singularity, corresponding to a period monodromy. Only two of these monodromies are independent, and in the following we will restrict to the large complex structure and the conifold monodromy. We can choose a period basis where the monodromy transformations are $\Pi \rightarrow {\bf T}_{i} \Pi$, with monodromy matrices~\cite{Denef:2001xn}
\begin{eqnarray}\label{eq:monmatrices}
{\bf T}_{LCS} = \left( \begin{array}{cccc}
1 & 1 & 3 & -5 \\
0 & 1 & -5 & -8 \\
0 & 0 & 1 & 1 \\
0 & 0 & 0 & 1 \end{array} \right) , \ \ \  
{\bf T}_{C} = \left( \begin{array}{cccc}
1 & 0 & 0 & 0 \\
0 & 1 & 0 & 0 \\
0 & 0 & 1 & 0 \\
1 & 0 & 0 & 1 \end{array} \right).
\end{eqnarray}

We now turn on three-flux, which will both lift the complex structure modulus and back-react on the geometry of the internal manifold. For small fluxes, the effect will be a warped internal manifold that is conformally Calabi--Yau. The moduli space of this manifold is similar to that of the original Calabi--Yau, but warping affects the metric of the moduli space~\cite{DeWolfe:2002nn}. To facilitate our analysis we assume that the overall-scale K{\"a}hler modulus, $\rho$, is fixed at large compactification volume. In general this is highly non-trivial (see~\cite{Denef:2008wq} for a clear description of some of the difficulties), but has been accomplished in a variety of scenarios~\cite{Kachru:2003aw,Balasubramanian:2005zx}. For such large-volume compactifications warping can be neglected~\cite{Giddings:2001yu}, although there are a number of important subtleties to address in the dimensional reduction~\cite{Giddings:2005ff,Shiu:2008ry,Douglas:2008jx}. We must also include D3-b
 ranes and O3 planes or D7-branes to cancel the overall D3 charge induced on the compact internal manifold by the flux~\cite{Giddings:2001yu}. We neglect the influence of these localized objects on the potential for the complex structure and axio-dilaton. 

With the caveat that we are working under the above stated set of assumptions, we can describe the qualitative features of a flux-induced potential on the axio-dilaton and complex structure moduli space of the Mirror Quintic, with the salient technical details relegated to Appendix~\ref{ApB}. The potential induced for the axio-dilaton, $\tau$, is particularly simple. At fixed $z$ for a given flux configuration, the extrema of the potential are determined by a quadratic equation in $\tau$~\cite{Danielsson:2006xw}, the physical root of which corresponds to a global minimum. The imaginary part of $\tau$ determines the string coupling, which we must ensure is less than unity in order to maintain the validity of the supergravity approximation. This will be explicitly determined in all of the examples we present.
 
We now describe the potential on complex structure moduli space. Type IIB string theory has two types of three-form fluxes, Ramond-Ramond (RR) and Neveu-Schwarz (NSNS), which can be arranged into vectors $F = (F_1, F_2 , F_3, F_4)$ and $H = (H_1 , H_2, H_3, H_4)$ respectively. In our conventions, $\Pi_1$ corresponds to the period of the shrinking cycle associated with the conifold, with $F_4$ and $H_4$ the flux piercing this cycle. $F_1$ and $H_1$ are the fluxes piercing the dual cycle $\Pi_4$. The flux induces a superpotential \cite{Gukov:1999ya}
\be
W = (F-\tau H) \cdot \Pi,
\ee
from which the $\cN=1$ scalar potential can be computed. For generic fluxes, there will be local minima of the potential, as we describe in more detail below. Since the monodromies introduced in Eq.~\ref{eq:monmatrices} preserve the symplectic structure, we can think of them as acting either on the period vectors {\em or equally well, as acting on the flux vectors}, yielding $F' = F {\bf T}$ and $H' = H {\bf T}$. This is in exact analogy to the heuristic example of fluxes on a torus introduced in Sec.~\ref{modstabland}. 

Although the monodromy transformations change the flux through the three-cycles, it is important to note that the transformations preserve the symplectic structure. By Gauss' law, the total charge on a compact manifold must always vanish. In the compactifications discussed here, this turns in to a tadpole condition on the fluxes:
\be
F \cdot Q \cdot H = N,
\ee
where $N$ is set by the total charge of branes and orientifold planes in the compactification, and $Q$ is the intersection matrix of three-cycles (see Eq.~\ref{eq.Q}). This symplectic flux product is unchanged by the monodromies, and there is no need to nucleate any branes when going from one minimum to another.

Note that in general, the monodromy transformations connect flux configurations that do not differ by one unit of flux; for example, we have $F' = (F_1 + F_4 , F_2, F_3, F_4 )$ under a conifold monodromy. We can always choose $H_4 = 0$ using the invariance of SL(2,Z) transformations of the axio-dilaton and fluxes in Type IIB~\cite{Danielsson:2006xw}, and will do so in what follows. Given a particular flux configuration, one cannot reach all other flux configurations by performing the monodromies under consideration,  leaving open the possibility that the landscape consists of "islands" of continuously connected sets of vacua (although it is possible to connect many more configurations by considering an extended moduli space, see Ref.~\cite{Chialva:2007sv}).

\subsection{The four-dimensional action}

\begin{figure*}
\begin{center}
\includegraphics[height=8cm]{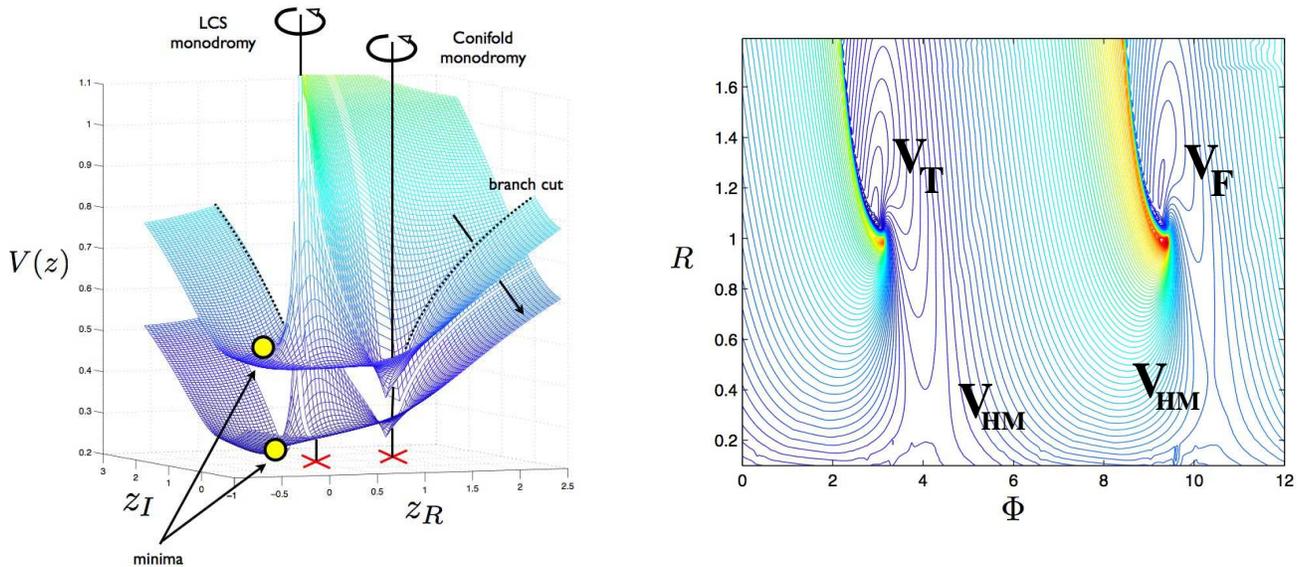}
\end{center}
\caption{The potential for the Mirror Quintic complex structure modulus is multivalued due to the three-cycle monodromies. The panels show two minima connected by a conifold monodromy. The upper minimum $V_F$ has flux configuration $F = [-2,-6,-9,-1]; H = [-1,0,-7,0]$ and the lower minimum $V_T$ has the flux configuration $F = [-3,-6,-9,-1]; H = [-1,0,-7,0]$. The polar coordinates used in the right panel are $R^2 = z_I^2 + (z_R- 1)^2$ and $\tan( \Phi) = z_I / (z_R - 1)$.}
\label{fig:sheets}
\end{figure*}

The four-dimensional effective action for this sample landscape can now be constructed. Collecting the results from Appendix~\ref{ApB}, the action is
\begin{eqnarray} \label{eq:4Daction}
S = \frac{M_p^2}{2} \int d^4x & \sqrt{-g}& ( R - 2 K_{\tau \bar{\tau}} \partial_{\mu}\tau \partial^{\mu}\bar{\tau} \\
  &-& 2 K_{z \bar{z}} \partial_{\mu}z \partial^{\mu}\bar{z} - \frac{2 M_p^2 g_s}{\pi \rho_I^3} v (\tau,z) ) \nonumber.
\end{eqnarray}
We will often find it convenient to write the action in terms of real fields $\phi^i \in \{z_R, z_I, \tau_R, \tau_I\}$, where $z = z_R + i z_I$. The non-zero components of the K{\"a}hler metric are then $K_{z_Rz_R} = K_{z_Iz_I} = K_{z\bar{z}}$ and $K_{\tau_R\tau_R} = K_{\tau_I\tau_I} = K_{\tau\bar{\tau}}$. The metric is, in the real field basis $\phi^i$,
\[
G_{ij} = 4\left( 
\begin{array}{cccc}
K_{z\bar{z}} & 0&0&0\\
0&K_{z\bar{z}} & 0&0\\
0&0& K_{\tau\bar{\tau}}&0\\
0&0&0& K_{\tau\bar{\tau}}\\
\end{array}
\right).
\]
Thus the kinetic terms are given by
\be
2 K_{\tau \bar{\tau}} \partial_{\mu}\tau \partial^{\mu}\bar{\tau}+ 2 K_{z \bar{z}} \partial_{\mu}z \partial^{\mu}\bar{z} = \frac{1}{2}G_{ij} \partial_{\mu}\phi^i  \partial^{\nu}\phi^j .
\ee

As discussed above, $\tau$ exhibits a global minimum for fixed $z$ and flux. Along some path in $z$, the position of this minimum will drift, but dynamically $\tau$ will typically remain in the neighborhood of the minimum. We will therefore be interested in taking a slice of the potential where $\tau$ is everywhere minimized, and will study the combination 
\begin{equation}
V_N \equiv g_s (\tau_{min}) v (z, \tau_{min}),
\end{equation}
unless otherwise noted.  

The periods of the four three-cycles of the Mirror Quintic are given by Meijer G-functions, as described in~\cite{Chialva:2007sv, Denef:2001xn}, allowing us to calculate $V_N$, $K_{\tau \bar{\tau}}$, and $K_{z \bar{z}}$ numerically. We compute the potential for each flux configuration on a square grid with cell size $\Delta z_{I,R} = .04$ using the built-in Meijer G-functions of Maple. Near the conifold point, the convergence of these numerical expressions is exceedingly slow, and we are unable to determine the scalar potential and K{\"a}hler potential. However, there do exist analytic expressions for the K{\"a}hler potential and period vectors in the very-near vicinity of the conifold point, see e.g.~\cite{Chialva:2007sv} and references therein.  

An example of the numerically generated potential is shown in Fig.~\ref{fig:sheets}. In the left panel, the multi-sheeted structure of the potential is manifest. Moving across the branch cuts emanating from the conifold and LCS points brings the field to another level of the potential, forming two spiral staircases. In this example, there are minima on both the upper and lower sheet that are continuously connected by performing a conifold monodromy. The upper minimum has the flux configuration $F = [-2,-6,-9,-1]; H = [-1,0,-7,0]$, and the lower minimum $F = [-3,-6,-9,-1]; H = [-1,0,-7,0]$. We can perform any number of conifold and LCS monodromies, creating a large number of levels, and potentially a large number of continuously connected minima~\cite{Danielsson:2006xw, Chialva:2007sv}. We will also find it convenient to define polar coordinates about the conifold point, $R^2 = z_I^2 + (z_R- 1)^2$ and $\tan( \Phi) = z_I / (z_R - 1)$, which allows us to "unwrap" the potential a
 nd remove one of the branch cuts. A contour plot of the potential in this representation is shown in the right panel of Fig.~\ref{fig:sheets}. The two minima are indicated by $V_T$ and $V_F$, with $V_F > V_T$. There are also two other extrema, labeled as $V_{HM}$, corresponding to saddles. 

We have generated potentials for a number of different flux configurations. Shown in Table~\ref{tab:examples} are some examples. Each numbered sequence (e.g. 1a) and 1b)) represent  potential sheets connected by a conifold monodromy. This can correspond to a change in flux by one unit or many units (one unit for examples 2,3,4, two units for examples 1 and 5, and six units for example 6), and there exist sequences of a number of connected minima, as in example 2. The minima have a variety of potential values, and in general, because the flux has a number of components, increasing the value of one of the components of the flux upon the application of a conifold monodromy does not necessarily increase the value of the potential. The eigenvalues $\lambda_{1,2}$ of the dimensionless mass matrix
\begin{equation}
m_{ij}^2 = 4 K_{z \bar z} \frac{\partial^2 V_N}{\phi^i \phi^j},
\end{equation}
evaluated at the extrema are in most examples order one, yielding a mass scale for the complex structure moduli of order $m_z^2 \simeq M_p^2 \rho_I^{-3}$. The minima typically are not symmetric, and have one large and one small eigenvalue. The additional extremum is a saddle point, with one positive and one negative eigenvalue. The negative eigenvalue is also typically of order $M_p^2 \rho_I^{-3}$. The size of $g_s= \tau_I^{-1}$ at its minimum is typically of the order 0.1.

Having discussed the potential and some of its properties, we now digress on its range of validity. As discussed in Section~\ref{sec:scales}, the effective field theory approach is self-consistent only when energies remain below the string and Kaluza-Klein scales. The ratio of the relevant scales is given by
\begin{eqnarray}
\frac{V}{M_s^4} = (\pi g_s)^{-1} V_N, \ \ \frac{V}{M_{KK}^4} = \pi^{-1} \rho_I V_N, \\
\frac{m_z}{M_{s}} \sim g_s^{-1/4} \rho_I^{-3/4}, \ \ \frac{m_z}{M_{KK}} \sim \rho_I^{-1/2},
\end{eqnarray}
where the factors of $g_s$ arises in the first two relations due to the fact that $V_N$ incorporates $g_s (\tau_{min})$. For most of the examples at hand, we will have a potential that is close to or above the string and Kaluza-Klein scales. We should therefore expect that the potential will receive corrections from the massive Kaluza-Klein modes~\footnote{The massive KK modes were included in the dimensional reduction of Ref.~\cite{Shiu:2008ry}; we leave an examination of their effects to future work.} and/or stringy degrees of freedom (from Eq.~\ref{eq:scales} the hierarchy between the string and KK scales is determined by the relative size of $g_s$ and $\rho_I$, and determines the relative importance of these two possible corrections). Nevertheless, we still expect the generic features of the potential, such as the existence of minima connected by monodromies, to survive. We will therefore proceed with caution, acknowledging that the potential between the minima and detail
 ed quantitative predictions of this model will most likely receive corrections in a more complete description. 

\begin{table*}
\begin{ruledtabular}
\begin{tabular}{l c c c c c c}
Flux Configuration: & $V_N^{\rm min}$ & $m_{11}^{\rm min}$ & $m_{22}^{\rm min}$ & $V_N^{HM}$ & $m_{11}^{HM}$ & $g_s $ \\
1a) $F=[3, -4, -1, -2];H=[1, 0, 5, 0]$ & 2.8 & 1.2 & .25  & 2.87 & -1.4 & .23 \\
1b) $F=[1, -4, -1, -2]; H=[1, 0, 5, 0]$ & 2.67 & 1.2 & .2 & 2.77 & -.75 & .27 \\
2a) $F=[-2, -6, -9, -1]; H=[-1, 0, -7, 0]$ & .24 & 2.4 & .6 & .32  & - 2.4 & .25 \\
2b) $F=[-1, -6, -9, -1]; H=[-1, 0, -7, 0]$ & .03 & 2.3 & .5 & .21 & -3 & .39 \\
2c) $F=[0, -6, -9, -1]; H=[-1, 0, -7, 0]$& 0 & 2 & .38 & .13 & -2.4 & .27 \\
2d) $F=[1, -6, -9, -1]; H=[-1, 0, -7, 0]$ & 0 & 1.8  & .7 & .073 & -2.1 & .25 \\
3a) $F = [2,9,-4,1]; H=[-1,0,-7,0]$& 8.66 & 5.2 & .5 & 8.67 & -1 & .14 \\
3b) $F = [1,9,-4,1]; H=[-1,0,-7,0]$ & 8.45 & 3.9 & .8 & 8.52 & -2 & .16 \\
4a) $F=[1, 9, 7, 1];H=[1, 0, 6, 0]$ & .27 & 3.2 & .7 & .37 & -3 & .15 \\
4b) $F=[2, 9, 7, 1];H=[1, 0, 6, 0]$ & .41 & 3.7 & .75 & .486 & -1.8 & .12  \\
5a) $F=[1, 2, -8, -2];H=[2, 2, 9, 0]$ & .644 & 2.6 &.75  & .716 & -3 & .65 \\
5b) $F=[-1, 2, -8, -2];H=[2, 2, 9, 0]$ & .638 & 2.8 & .8 & .710 & -3.8 & .65 \\
6a) $F=[-6, -3, 5, 6];H=[1, 1, 1, 0]$ & 1.62 & 2.4 & .8 & 1.687 & -2 & .58 \\
6b) $F=[0, -3, 5, 6];H=[1, 1, 1, 0]$ & 1.56 & 2 & 1&  &  & .6 \\
 \end{tabular}
 \end{ruledtabular}
 \begin{center}
 \caption{Properties of some example flux-induced potentials. \label{tab:examples}}
 \end{center}
\end{table*}

\section{Dynamics of the moduli}
\label{sec:dyn_mod}

Cosmological solutions in the presence of the action Eq.~\ref{eq:4Daction} can be determined by assuming a FRW metric
\begin{equation}
ds^2 = -dT^2 + R^2 \left[ \frac{d \bar{r}^2 }{1-k \bar{r}^2} + \bar{r}^2 d \Omega_2^2 \right],
\end{equation}
where $k = \{-1,0,1\}$ for an open, flat, or closed universe respectively. In general, it is possible to decompose a scalar potential as $V = \mu^4 V_N$, where from Eq.~\ref{eq:4Daction}, the flux-induced potentials will have $\mu^4 \equiv M_p^4 (\pi \rho_I^3 )^{-1}$. Expressing the action in terms of the real fields $\phi^i$, and defining $M^2 = M_p^2/2$ and the dimensionless coordinates
\begin{equation}
t = \frac{\mu^2}{M} T, \ \ r = \frac{\mu^2}{M} R,
\end{equation}
the action then becomes
\begin{equation}\label{eq:frwaction}
S_E = 4 \pi^2 \frac{M^4}{\mu^4} \int dt r^3 \left( \frac{\epsilon^2 }{r^2}(k -\dot{r}^2) + \frac{1}{2}G^{ij} \dot{\phi}_i \dot{\phi}_j  - V_N(\phi) \right),
\end{equation}
where $\epsilon^2$ is defined as
\begin{equation}
\epsilon^2 \equiv \frac{M^2}{3 M_p^2}.
\end{equation}

The equations of motion are given by
\be
\begin{split}
&\dot{r}^2 = -k + \epsilon^2 r^2 \left(\frac{1}{2}G^{ij}\dot{\phi}_i\dot{\phi}_j + V_N \right), \\
&\ddot{\phi}^i + \frac{ 3\dot{r} } { r } \dot{\phi}^i + \Gamma^i_{jk} \dot{\phi}^j\dot{\phi}^k
 = - G^{ij}\frac{\partial V_N}{\partial \phi^j},
\end{split}
\label{eq:EOM2}
\ee
where $\Gamma^i_{jk}$ are the Christoffel symbols on field space. 

\subsection{Inflation}

Cosmologically, perhaps the most interesting dynamical behavior we could hope to find in our sample landscape is inflation. We will only briefly discuss this here, and refer the reader to recent reviews on inflation in string theory~\cite{McAllister:2007bg,Burgess:2007pz} for more details. To determine if a potential is suitable to drive an epoch of inflation, we must satisfy the slow-roll conditions
\begin{eqnarray}
\frac{1}{2 V^2} G^{i j} \frac{\partial V}{\partial \phi^i} \frac{\partial V}{\partial \phi^j} \ll 1, \\ 
{\rm min} (\lambda) \ll 1,\\
\end{eqnarray}
where $i$ and $j$ run over all the dimensionless moduli fields $\phi^i$, and ${\rm min} (\lambda)$ is the smallest eigenvalue of the matrix
\begin{equation}
N^i_j = \frac{1}{V} G^{ik} \left( \frac{\partial^2 V}{\partial \phi^j \partial \phi^k} - \Gamma^{l}_{jk} \frac{\partial V}{\partial \phi^l} \right).
\end{equation}
Note that the overall scale of the potential is irrelevant in determining the slow roll parameters. There are two important obstacles to finding inflation in multiple field models of supergravity. First, we must ensure that the gradient of the potential is small {\em in all directions of field space} in order to satisfy the first slow-roll condition. We must also ensure that there are no corrections to the potential that will spoil slow-roll. This problem is particularly acute in models of supergravity, where corrections to the K{\"a}hler potential can cause a so-called $\eta-$problem, rendering the second derivative too large to support significant (or any) inflation~\cite{Copeland:1994vg}. Most models of stringy inflation fall prey to one of these two problems (for a manifestation of the $\eta-$problem, see eg~\cite{Baumann:2007np}).

In our sample landscape, we have assumed that $\rho_I$ is fixed, and we neglect the other K{\"a}hler moduli, potentially introducing the first problem. In addition, both the superpotential and the K{\"a}hler potential are corrected beyond the classical approximation, potentially introducing the second problem. Nevertheless, we can check if the potential over the complex structure moduli space as computed here is at least consistent with driving an epoch of slow-roll. To do so, we have computed the slow-roll parameters for a variety of flux configurations. In most situations, the first slow-roll parameter is much larger than one except for the near-vicinity of the saddle. This is mainly a consequence of the fact that $G^{i j} \gg 1$ over most of moduli space. Very near the LCS and conifold points $G^{i j}$ can become small, but this is where the potential drastically steepens, as can be seen in
Fig.~\ref{fig:sheets}. In the LCS region there are polynomial expansions for the periods (see~\cite{Denef:2001xn} for the Mirror Quintic expressions), and using these it can be shown that the first slow-roll parameter approaches a  $\mathcal{O}(1)$ constant. Also, the analytical expansion near the conifold point (see~\cite{Danielsson:2006xw}) can be shown to violate the first slow-roll condition.  

The only potentially viable location for slow roll inflation is therefore in the vicinity of the saddle point. However, here the relevant eigenvalue of the mass matrix is typically not small enough to ensure that the second slow-roll condition is satisfied. We have found examples where a minimum can nearly merge with the saddle to produce an approximate inflection point. This will produce a region of the potential where both slow-roll conditions are satisfied over a small range, and falls under the rubric of accidental inflation~\cite{Linde:2007jn}. However, we stress that this scenario is extremely vulnerable to corrections and may simply fall apart when the neglected K{\"a}hler moduli are re-introduced. 

We therefore conclude that aside from some potential very fine-tuned scenarios, it is difficult to realize inflation in our sample landscape. This is likely to carry over to flux potentials on other compactification manifolds.

\subsection{General trajectories}\label{sec:gentraj}
In general, the cosmological dynamics of moduli fields can be quite complicated. The equations of motion are non-linear and trajectories can exhibit chaotic behavior. Fortunately, in our example the behavior of trajectories on moduli space is rather simple. As described above, the potential for $\tau$ always has a global minimum at fixed $z$, and we will assume initial conditions for which $\tau$ remains near this minimum.  Therefore, we will restrict ourselves to motion along paths in the four-dimensional $z-\tau-$space where $\partial_{\tau}V=0$.

\begin{figure}[tb]
\begin{center}
\includegraphics[height=7.3cm]{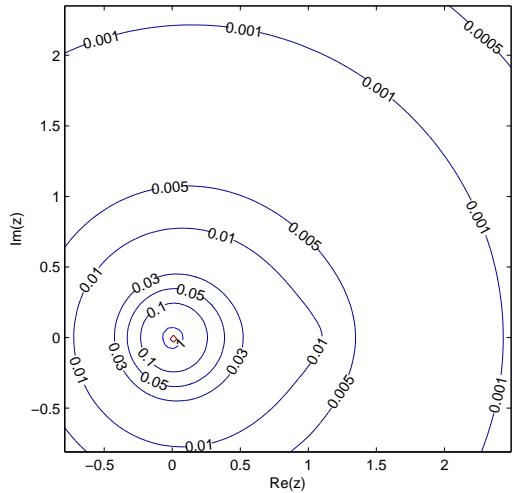}
\end{center}
\caption{The metric $K_{z \bar{z}}$ on the complex $z$-plane is almost rotationally symmetric around the LCS point. The metric diverges near the LCS point $z=0$ and near $z=1$ where  $K_{z \bar{z}} \sim ln|z-1|$.}
\label{fig:kzz}
\end{figure}

Considering trajectories in $z-$space, we must first describe the behavior of $K_{z {\bar z}}$. The K{\"a}hler potential is independent of flux~\footnote{The flux-induced warping of the internal manifold yields flux-dependent corrections to the K\"ahler potential, which we ignore.}, and so will be identical for each flux configuration. The metric coefficient $K_{z \bar{z}}$ diverges as $z \rightarrow 0$, and goes to zero as $|z| \rightarrow \infty$  as shown in the contour plot Fig.~\ref{fig:kzz}. As seen from the figure, aside from the near-conifold region, $K_{z \bar{z}}$ depends only on $|z|$ to first approximation. Outside the near-neighborhood of the LCS and conifold singularities it falls off approximately like $|z|^{-2}$.

\begin{figure*}[tb]
\begin{center}
\includegraphics[height=6.3cm]{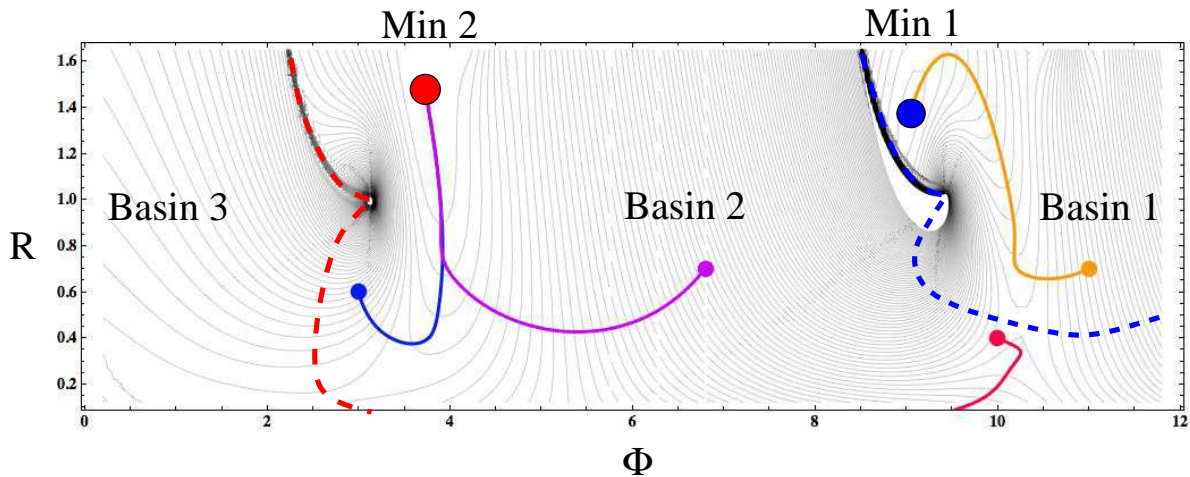}
\end{center}
\caption{Here, we show the potential corresponding to a flux configuration in the upper minimum (Min 1, indicated by the filled circle to the right) given by $F=[3, -4, -1, -2]; H=[1, 0, 5, 0]$ (example 1 in Table~\ref{tab:examples}). The lower minimum, Min 2, is denoted by the filled circle on the left. Some sample trajectories are shown as solid lines, with initial positions denoted by small filled circles (the initial velocity is zero in these examples). The numerical  evolution cannot be tracked into the near-conifold region, where the trajectory on the bottom right disappears. The potential splits into three basins of attraction, whose boundaries are denoted by the dashed lines. Zero-velocity initial conditions in Basin 1 will yield trajectories that end in Min 1, trajectories that begin in Basin 2 will end in Min 2, and trajectories that begin in Basin 3 will fall to the next sheet down, possibly reaching a minimum should one exist. }
\label{fig:basins}
\end{figure*}

We have solved the equations of motion Eq.~\ref{eq:EOM2} for a flat FRW universe in the presence of a number of flux-induced potentials for a variety of initial conditions using an adapted version of the Supercosmology program~\cite{Kallosh:2004rs}. Some example trajectories are plotted in Fig.~\ref{fig:basins}. Because determining the potential near the conifold point is computationally expensive, we will not be able to track the evolution of the fields with good accuracy in this region (the potential becomes fairly inaccurate at a distance $|z-1| \sim .1$). Away from the conifold point, we define an interpolating function over the grid of data for the two-dimensional potential. This introduces a small numerical error in determining the gradient of the potential at each step of the numerical integration, but we do not expect this to significantly alter the qualitative behavior of the computed trajectories.  

Considering initial conditions with field velocities that are not significantly larger than order one, trajectories that do not pass too close to the conifold point (where the numerics break down) will end up either in the minimum on the sheet where the initial conditions were defined, or will pass around the LCS or conifold point to the next sheet down. If there is another minimum on the lower sheet, the trajectory will typically end here. By considering a number of trajectories with zero initial velocity, we have mapped out the approximate basins of attraction for the two adjacent minima shown in Fig.~\ref{fig:basins}. Initial conditions in Basin 1 will result in trajectories that end in Min 1, trajectories beginning in Basin 2 will end in Min 2, and trajectories beginning in Basin 3 will fall to the left, potentially into another minimum on the next sheet down. As expected, the saddle points lie on the boundaries between basins. Note also that the lower saddle (in the vici
 nity of Min 2) is closer to the conifold point than the upper saddle (in the vicinity of Min 1). This seems to be a generic property of the examples we have studied, and accords with our observation that trajectories typically find the nearest minimum as opposed to spanning many sheets of the potential.

The potential diverges at the conifold point if there is flux through the shrinking cycle ($\Pi_1$, in our example)~\footnote{In the case where there is no flux through the shrinking cycle, the conifold point is a minimum of the potential. Thus, it seems that such a flux configuration would enforce a geometric transition of the internal manifold, as discussed in~\cite{Danielsson:2006xw, Chialva:2007sv}.}. We can therefore be assured that those trajectories that we cannot track numerically eventually leave the near-conifold region. 

\section{Instanton calculations of tunneling rates}
\label{sec:tunneling} 

The moduli parametrizing the the string theory landscape can be frozen at the local minima of a multi-dimensional potential as discussed above. Classically, this configuration is stable, and will correspond to a four-dimensional Minkowski, de Sitter, or anti-de Sitter vacuum. Semi-classically, however, many of these vacua will be unstable, tunneling to a region of the potential with lower energy density. This process, originally described by Coleman and collaborators~\cite{Coleman:1977py,Callan:1977pt,Coleman:1980aw}, will proceed through the nucleation of bubbles of a lower energy phase that then expand, eating up the original vacuum. The tunneling rate out of the false vacuum is in the WKB approximation~\cite{Coleman:1980aw} given by
\be
\Gamma \simeq A e^{-(S_{E} [g_E,\phi] - S_{E} [g_E,\phi_F])},
\label{eq:tunrate}
\ee
where $S_E$ and $g_E$ are the Euclidean continuation of the action $S$ and the metric $g$ and $A$ is a pre-factor representing the first quantum corrections to the rate~\footnote{The pre-factor in Eq~\ref{eq:tunrate} involves a functional determinant that has only been calculated in the absence of gravitational effects~\cite{Callan:1977pt}.}. The background subtraction term in the exponential, $S_{BG} \equiv S_{E} [g_E,\phi_F]$, for a de Sitter false vacuum is given by
\begin{equation}
S_{BG} = - \frac{24 \pi^2  M_p^4}{V_F}.
\end{equation}

The instanton action, $S_I = S_E [g_E,\phi]$, must be determined by solving the Euclidean equations of motion, which for a general metric is a formidable task. However, there is evidence that the instantons yielding the largest tunneling rates (lowest action) are $O(4)$ invariant (this has been proven for tunneling in flat space~\cite{Coleman:1977th}). This means that the Euclidean metric has the form
\be
ds^2_E = \frac{M^2}{\mu^4} \left( d^2 \chi + r^2(\chi) d \Omega_3^2 \right),
\label{eq:FRWmetric}
\ee
and the field, $\phi = \phi(\chi)$, is a function only of $\chi$. The equations of motion follow from the Euclideanized version of the action Eq.~\ref{eq:frwaction} with $k=1$ and $\chi = i t$
\be
\begin{split}
&\dot{r}^2 = 1 + \epsilon^2 r^2  
\left(\frac{1}{2}G^{ij}\dot{\phi}_i\dot{\phi}_j - V_N \right), \\
&\ddot{\phi}^i + \frac{ 3\dot{r} } { r } \dot{\phi}^i + \Gamma^i_{jk} \dot{\phi}^j\dot{\phi}^k
 = G^{ij}\frac{\partial V_N}{\partial \phi^j},
\end{split}
\label{eq:EOMEuc}
\ee
where $\Gamma^i_{jk}$ are the Christoffel symbols on field space. We remind the reader that all variables appearing in these equations of motion are dimensionless and real. When the false vacuum energy is not zero the Euclidean manifold will be compact, and the equations of motion will be subject to the boundary conditions
\be
\begin{split}
r(\chi = 0) = 0 \mbox{ , } \dot{\phi}(\chi = 0) = 0 \mbox{ , } \phi(\chi = 0) \sim \phi_F \\
r(\chi = \chi_{max}) = 0 \mbox{ , } \dot{\phi}(\chi = \chi_{max}) = 0.
\end{split}
\label{eq:BC1}
\ee
Thus, all coordinates of the instanton metric have a finite range, and the action is finite. Note that we do not require that the field ends up at a lower energy minimum. Instead, the field will emerge at some point on the other side of a potential barrier, perhaps in the basin of attraction of some lower energy (true vacuum) minimum, and subsequently roll classically toward it. Using the equations of motion, the instanton action is found by evaluating
\begin{equation}
\label{eq:S_I}
S_I = 4 \pi^2 \frac{M^4}{\mu^4} \int d\chi \left( r^3 V_N - \frac{r}{\epsilon^2} \right),
\end{equation}
over the Euclidean manifold.

\subsection{One scalar field coupled to gravity}\label{sec:onescalar}
\begin{figure}
\begin{center}
\includegraphics[height=4cm]{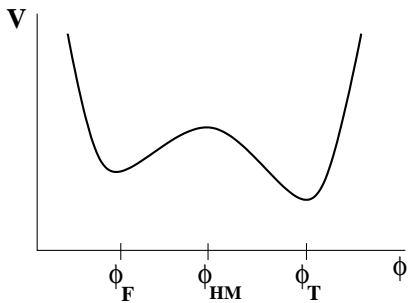}
\end{center}
\caption{A potential for a field $\phi$  with two minima and one maximum.}
\label{fig:1dpot}
\end{figure}

Before treating the general problem of multiple scalars, let us review the calculation of the instanton for one scalar coupled to gravity. Consider a single scalar with a potential exhibiting two local minima, as shown in Fig.~\ref{fig:1dpot}. A field redefinition gets rid of the metric in the kinetic term of the action Eq.~\ref{eq:frwaction}, and the Euclidean equations of motion for the field reduce to
\be
\begin{split}
&\dot{r}^2 = 1 + \epsilon^2 r^2  
\left(\frac{1}{2}\dot{\phi}^2 - V_N \right), \\
&\ddot{\phi} + \frac{ 3\dot{r} } { r } \dot{\phi} 
 = \frac{\partial V_N}{\partial \phi}.
\end{split}
\label{eq:EOM1}
\ee
 
The number of possible instantons has to be determined on a case-by-case basis. However, gravitational effects guarantee the existence of one instanton, the Hawking--Moss (HM) instanton \cite{Hawking:1981fz}. Note that the potential has a local maximum $V_N (\phi_{HM})$ in between the minima (see Fig.~\ref{fig:1dpot}). Thus there is a trivial solution to Eq.~\ref{eq:EOM1} and Eq.~\ref{eq:BC1} where the field $\phi$ is constant and $r$ is compact:
\be
\phi = \phi_{HM}, \ \  r = (\epsilon V_N^{1/2})^{-1} \sin (\epsilon V_N^{1/2} \chi).
\ee
The field can fluctuate from the false vacuum to this maximum with probability
\begin{eqnarray}
\Gamma &\sim& e^{-(S_I - S_{BG})} \nonumber \\ 
&\sim& exp \left[ \frac{M^4}{\mu^4} \frac{8 \pi^2}{3 \epsilon^2}  \left( \frac{1}{V_N(\phi_{HM})}- \frac{1}{V_N(\phi_{F})}\right) \right].
\label{eq:HM}
\end{eqnarray}
This is made explicit by studying the stochastic evolution of a field in de Sitter space, where the probability to random-walk up the potential matches Eq.~\ref{eq:HM}~\cite{Starobinsky:1986fx}. Since the field can subsequently roll into either of the two minima, this instanton will contribute to the tunneling rate out of the false minimum. 

If other instantons exist, they can be characterized by the number of times the field passes the local maximum of the potential during the instanton evolution. Note that the Euclidean equations of motion for $\phi$ in Eq.~\ref{eq:EOM1} are equivalent to the classical equation of motion of a particle moving in an upside-down potential. A particle rolling in this potential can either pass the local minimum once or oscillate around it several times. The former case is referred to as a single--pass instanton, while the latter are called multiple--pass instantons. In either case, actually computing the exact instanton by solving the equations of motion is a numerically challenging task. However, their existence can be proven by the overshoot/undershoot argument of Coleman~\cite{Coleman:1977py}. For the readers' convenience, we repeat the argument here. 

Consider the particle analogue mentioned above. The particle in the upside-down potential is subject to a friction ($\frac{ 3\dot{r} } { r} < 0$) or antifriction ($\frac{ 3\dot{r} } { r } > 0$) force. For the compact Euclidean metrics studied here, we have friction (antifriction) during the beginning (end) of the evolution of the field.  If the particle starts at rest too far from the local maximum at $\phi_F$, it will reach a turning point far before $r=0$; it will roll back toward $\phi_F$. The antifriction term will then push $\phi \rightarrow -\infty$ yielding a divergent instanton action. If the particle instead starts too close to $\phi_F$, it will have enough energy, with the aid of the antifriction force, to overshoot the maximum at $\phi_T$, thus pushing $\phi \rightarrow \infty$, yielding a divergent action. By continuity of the potential, there must be an intermediate point where the particle has just enough energy so that, taking the friction and antifriction into
  account, it reaches the end of its trajectory with zero velocity at the moment $r = 0$. Thus the single--pass instanton only exist for suitable friction/antifriction force terms, which depend upon the geometry of the instanton. Similar arguments can be applied to multiple-pass instantons~\cite{Hackworth:2004xb,Banks:2002nm}.

For any pair of minima, there will be a finite number of instantons mediating the transition~\cite{Hackworth:2004xb}~\footnote{Counting the negative modes, it is unclear that the multiple-pass instanton really represents a decay process~\cite{Batra:2006rz,Lavrelashvili:2006cv,Gratton:2000fj}.}. However, in any conceivable situation, one instanton will give the lowest action, and dominate the tunneling process. The HM instanton is the dominant decay channel when the barrier in between the minima is broad, making gravitational effects important. Qualitatively this can be understood by the following argument. Sending $\epsilon \rightarrow \infty$  (that is, considering a family potentials related by increasing the ratio of the width to the height, or equivalently, considering the strong--gravity regime; see~\cite{Aguirre:2006ap} for a similar construction) implies, by Eq.~\ref{eq:EOM1}, that $\rho \rightarrow 0$ and the range of $\phi$ is limited to the constant value $\phi_{HM}
 $. Therefore, the only possible instanton is the HM instanton, and all other instantons that existed for finite $\epsilon$ will shrink to this point. Quantitatively, a more stringent analysis shows that the HM instanton is the only possible solution to Eq.~\ref{eq:EOM1} if $d^2 V(\phi_{HM}) / d\phi^2 < 4 \epsilon^2 V_N (\phi_{HM})$~\cite{Hackworth:2004xb}. If instead $d^2V(\phi_{HM})/d\phi^2 > 4 \epsilon^2 V_N (\phi_{HM})$ it is the single-pass, or Coleman--de Luccia (CDL), instanton that yields the largest tunneling rate. Thus, when the barrier in between the minima is narrow compared to its height, the semi-classical tunneling dominates the stochastic fluctuations  described by the HM instanton. 

\subsection{The thin-wall approximation}\label{sec:thinwall}
When the energy splitting between vacua becomes much smaller than the height of the barrier separating them, then the field must loiter in the neighborhood of each vacuum for the majority of the Euclidean evolution, only quickly making the transition from one vacuum to the other. The instanton can then be approximated as two de Sitter four-spheres joined across a thin interface--this is known as the thin-wall approximation. In this limit, the position of the wall becomes a collective coordinate, and the equations of motion for the instanton can be solved exactly. We can define the tension of the wall as
\begin{equation}
\label{eq:tension}
\sigma = M \mu^2 \int_{\phi_F}^{\phi_T} d\phi \sqrt{2 \left( V_0 (\phi) - V_0 (\phi_F)\right)} \equiv M \mu^2 \sigma_N,
\end{equation}
where $V_0$ is a (dimensionless) function that is identical to $V_N$ except in the vicinity of $\phi_F$, where the potential is deformed such that $V_0 (\phi_T) = V_0 (\phi_F)$ and $dV_0 / d\phi (\phi_{T,F}) = 0$~\cite{Coleman:1980aw}. Given this definition and the energy density in the true and false vacua, the (dimensionless) initial radius and Euclidean action of the bubble including gravitational effects is given by~\cite{Parke:1982pm}
\begin{equation}
r^2 = \frac{r_0^2}{1 + 2 x y + x^2},
\end{equation}
where
\begin{eqnarray}
r_0 &=& \frac{\sigma_N}{V_{N} (\phi_F) - V_{N} (\phi_T)}, \nonumber \\ 
x &=& \frac{3 \epsilon^2}{2\left( V_N (\phi_F) - V_N (\phi_T) \right)} , \nonumber \\ 
y &=& \frac{V_N (\phi_F) + V_N (\phi_T)}{V_N (\phi_F) - V_N (\phi_T)}.
\end{eqnarray}
The bounce action is given by
\begin{equation}\label{eq:thinwallaction}
S_I - S_{BG} = \frac{M^4}{\mu^4} \frac{27 \pi^2 \sigma_N^4 }{2 \left( V_N (\phi_F) - V_N (\phi_T) \right)^3} f(x,y),
\end{equation}
where
\begin{equation}
f(x,y) = 2 \frac{(1+xy) - (1+2 x y + x^2)^{1/2}}{x^2 (y^2 - 1) (1+2 x y+x^2)^{1/2}}.
\end{equation}

\subsection{Multiple scalar fields coupled to gravity}\label{sec:multscalars}

We now turn to the case of interest in the string theory landscape, where multiple scalar fields (moduli) are stabilized by a potential with many minima. We must now re-introduce the metric on field space, $G_{ij}$~\footnote{In the one-dimensional case, a field redefinition got rid of the metric in the kinetic term. In a multi-dimensional case, this is only possible in local patches of field space, not globally.}. In the one-dimensional case, the overshoot/undershoot argument could be used to prove the existence of instantons. A finite family of solutions was associated with any pair of minima, and each such family always contained the trivial HM solution. It is natural to expect that the same type of solutions exist in a multidimensional scenario. However, proving their existence  becomes much more difficult. Naively, we have an infinite number of paths in field space that would all need to be tested, which is clearly impossible. 

The multidimensional tunneling problem has been discussed previously by several authors (see e.g.~\cite{Banks:2002nm,Dasgupta:1996qu,Kusenko:1995jv,Moreno:1998bq,Konstandin:2006nd,Podolsky:2008du}). If gravity can be ignored and the field space metric is trivial ($G_{ij} = \delta_{ij}$) there are even numerical methods for computing the instantons~\cite{Kusenko:1995jv, Moreno:1998bq,Konstandin:2006nd}. However, it is not straight-forward to generalize these methods to the problem of moduli stabilization, where the metric on field space can be involved, and the gravitational effects from de Sitter space are inevitable. What we will do here instead is to discuss the qualitative features of multidimensional tunneling. We will identify situations where we can uniquely determine the path through field space, or at least some of its qualitative features, in which case the problem becomes effectively one-dimensional.

\subsubsection{Hawking--Moss instantons}
We begin by searching for the trivial solutions to the Euclidean equations of motion Eq.~\ref{eq:EOMEuc} satisfying the appropriate boundary conditions. These are the HM instantons, which sit at extremal points of the multidimensional potential while $r$ goes between its two zeros, and will exist as long as gravitational effects are non-negligeable. A multidimensional potential can have a diversity of extremal points, which we can classify by the number of negative eigenvalues of the matrix of second derivatives $\partial^2 V_N /\partial \phi^i \partial \phi^j$ evaluated at the extremal point. If there is at least one negative eigenvalue~\footnote{When all eigenvalues are positive or zero, the instanton cannot be interpreted as mediating the transition between two vacua--it will correspond to a Euclideanized vacuum solution or to an instanton that connects a vacuum to itself via an inflection point. The latter may be of cosmological interest, although the probability for such
  a transition will be low due to the large background subtraction.}, this unstable direction can connect the basins of attraction of two minima, and there should exist a HM instanton in complete analogy with the one-dimensional case. If there is more than one negative eigenvalue, then it is possible that the same HM instanton could mediate the transition between many different minima. Due to the background subtraction term, the transition probability to the HM point will not be identical if the initial minima have different cosmological constants. After the transition, the field will fall in any one of the (possibly numerous) unstable directions with equal probability. This implies that all vacua adjacent to a given HM point will be connected by such transitions.

In a one-dimension potential, the HM instanton reflected the inevitable semi-classical gravitational instability of a positive energy vacuum. However, in a multidimensional potential landscape, one can imagine scenarios where there are two vacua separated by a barrier that have no intervening extremal points. This could occur, for example, when there is an orthogonal direction in field space that has a non-zero gradient at every point between the vacua. Here, no trivial $O(4)$-invariant Euclidean solution exists, but the stochastic picture of the HM instanton suggests that an instability should still be present. This can be illustrated by considering a near-extremal point (in the sense that there might exist directions in field space in which first derivatives are small) between two vacua. The random walk of the field will be slightly biased by the gradient, but nevertheless could mediate a transition out of the original vacuum in the vicinity of the near-extremal point. If t
 he gradient is everywhere large, then a full stochastic analysis would be necessary to determine the semi-classical behavior of the field.  

\subsubsection{Single and multiple pass instantons}
We now consider non-trivial $O(4)$-invariant solutions to the Euclidean equations of motion. As mentioned previously, it is in general prohibitively difficult to find such solutions due to the multitude of possible paths, the importance of gravity, and the geometry over field space. However, progress can be made when the full potential has symmetries that determine the path that the instanton must take. For example, a deep valley typically must be followed in order to avoid singular solutions (since this becomes a precipitous ridge in the Euclidean potential). It is also possible to overcome these difficulties by concentrating on the evolution near an extremal point, where the easily--determined symmetries of the potential dictate the possible trajectories and the geometry on the field space is locally flat. This will be relevant when at least one of the instanton endpoints lies in the near-neighborhood of an extremal point (either a minimum or a HM point). We will make a num
 ber of precise statements in this context, that will be useful for determining important classes of instantons in multidimensional problems. Without such simplifications, the construction of the entire instanton solution will necessarily retain a qualitative flavor.

By concentrating on a local patch of the potential in the vicinity of an extremal point, we can make two important simplifications. If the patch is sufficiently small (as determined by the size of derivatives of the metric over field space), we can introduce locally inertial coordinates and neglect the geometry on field space. We can also Taylor expand the potential about the extremal point, which in terms of the locally inertial coordinates $x$ yields
\begin{equation}
V_N (\phi^i) \simeq V_0 + \frac{1}{2} m_{ij} x^i x^j,
\end{equation}
where $m_{ij} \equiv \partial^2 V_N / {\partial x^i \partial x^j} = G_{ii}^{-1/2} G_{jj}^{-1/2} \partial^2 V_N / {\partial \phi^i \partial \phi^j}$ is assumed to be non-zero.

The symmetries of the potential $V_N (\phi)$ in the neighborhood of the extremal point will depend entirely on the structure of $m_{ij}$. There will be continuous $O(n)$ symmetries when there are $n>1$ identical  eigenvalues of $m_{ij}$ (if there is more than one set of identical eigenvalues, there will be a product structure $O(n_j) \times O(n_{j-1}) \times \ldots O(n_1)$, where $n_j$ is the number of elements of each set). There will also be discrete reflection symmetries for each of the $d$ non-identical eigenvalues. If the potential does not admit a Taylor expansion, or if the lowest order term is not quadratic, then it will be possible to have other discrete symmetries.

Now, consider trajectories approaching an extremal point. Once more, it is helpful to consider the mechanical analogue of the tunneling problem, i.e. the particle rolling in the up-side down potential. The angular momentum with respect to the extremal point must be zero in order for the particle to reach it. The only zero angular momentum trajectories lie in one of the planes preserving an $O(n)$ symmetry or along one of the directions in field space preserving a discrete symmetry. We will refer to this class of trajectories as lying along a line of symmetry.

When the energy of a false vacuum is exactly zero (or when gravitational effects are negligible), then one of the instanton endpoints must lie {\em exactly} at the false vacuum minimum. The trajectory must therefore follow a line of symmetry in the neighborhood of the extremal point. When there are continuous symmetries, there are an infinite number of lines of symmetry, and there could be an infinite number of Euclidean trajectories that have an endpoint at the false vacuum. Of course, not all (or any for that matter!) of these trajectories need be instantons once we stray from our patch, but if the symmetry of the potential extends out from the false vacuum minimum, there can in fact exist an infinite family of instantons, and naively it seems as though the transition rate should go to one (since any one of this infinite number of transitions could carry one out of the false vacuum). However~\cite{Kusenko:1995bw,Kusenko:1996bv}, one must include these internal degrees of fr
 eedom in the calculation of the pre-factor in Eq.~\eqref{eq:tunrate}, which renders the overall probability finite, but enhanced by a factor of $(S_I - S_{BG})^{N/2}$, where $N$ is the number of continuous symmetries of the false vacuum broken by the transition. We leave the implications of such enhancements to future work.

When there are only discrete symmetries, the path of the instanton is constrained to lie along one of the (finite number of) lines of symmetry, and one has {\em directional} information about the velocity on the boundary of the coordinate patch. Inside the patch, this specifies a unique trajectory unstable to small perturbations, and implies that the number of instantons with an endpoint at the false vacuum is bounded by twice the number of discrete symmetries (twice because the instanton could potentially approach the false vacuum from either side of the extremal point). 

We now consider trajectories in the neighborhood of a HM point. As in the one-dimensional case, consider a family of potentials differing only by the value of $\epsilon$. In the limit where $\epsilon \rightarrow \infty$, only the HM instantons will exist. Decreasing $\epsilon$, eventually the range in Euclidean time will grow large enough to admit a single--pass CDL instanton. The endpoints of this instanton will necessarily lie in the near--neighborhood of the HM point.  In this case, the entire trajectory will be contained within our locally inertial patch, and we can use the simplified picture presented above to construct the instanton.

>From the boundary conditions, the initial and final angular momentum with respect to the HM point must be zero. When the mass matrix contains many different eigenvalues, the angular momentum with respect to the extremal point is not in general conserved. A trajectory beginning at a generic point with zero angular momentum will not evolve to a final state with zero angular momentum (at least not in our patch). Instead, the particle will orbit the HM point until the trajectory becomes singular due to the evolution of the scale factor. Therefore, we conclude that the instanton must lie along a line of symmetry (passing through the HM point), where the angular momentum can remain zero, and the problem becomes effectively one-dimensional~\footnote{It makes little sense to interpret instantons in directions associated with a continuous symmetry as tunneling solutions since there will be a classically allowed path between the instanton endpoints. Discrete lines of symmetry associate
 d with positive eigenvalues of the mass matrix should be ignored as well.}. Slicing the potential along one of the discrete lines of symmetry, we can determine the condition for the existence of a single--pass CDL instanton as
\begin{equation}\label{eq-instantonbound}
|\lambda_i| > 4 \epsilon^2 V_N (\phi_{HM}),
\end{equation}
where $\lambda_i$ is the appropriate eigenvalue of the mass matrix corresponding to the discrete line of symmetry of interest. Thus, there will be a class of instantons that behave as in the one-dimensional case; the multiple pass and the CDL instantons can be continuously deformed into the HM instanton as the relative importance of gravity increases. This argument does not rule out other classes of instantons in the weak gravity limit. However, we expect the class of instantons associated with an HM point to be relevant for the string theory landscape, where gravity is important.

\section{Tunneling in the Mirror Quintic monodromy staircase}
\label{sec:flux_tun}

\subsection{Euclidean paths}\label{sec:Eucpaths}
We now move on to the description of instantons mediating the transition between flux vacua in our Mirror Quintic sample landscape. We will have to contend with Euclidean evolution in the two complex dimensional $\tau-z-$space, a non-trivial K{\"a}hler metric, and gravitational effects, a set of ingredients that in general makes the problem untractable as described in Sec.~\ref{sec:multscalars}. However, we can use the intuition developed in previous sections and some simple features of the potential to help us find approximate solutions. 

Recall that minima on different potential sheets are connected by encircling the large complex structure or conifold point, as depicted in Fig.~\ref{fig:sheets}. In general, $\tau$ is not altered much between minima on adjacent sheets of the potential (see Table~\ref{tab:examples}). Given a particular value of $z$, the potential on the $\tau-$plane has a global minimum as described in Sec.~\ref{sec:MQ}. When we are looking for instantons, we are working with the upside-down potential, and this minimum is now a global maximum. We must be careful to get non-singular paths, and therefore $\tau$ must stay near this maximum as $z$ evolves. This renders the dynamics in $\tau-$space trivial, and we will (as before) restrict ourselves to motion on a slice where $\partial_{\tau} V = 0$.

As discussed in Section~\ref{sec:dyn_mod}, the K{\"a}hler metric $K_{z \bar{z}}$ takes a fairly simple form, and in general the Lorentzian trajectories are determined from the topography of the potential, and not strongly influenced by the geometry on field space. We should therefore expect that the Euclidean evolution of the field is also fairly well-behaved. To get a sense for the path length between two minima, one can fix the trajectory and define the canonical field along it. Specifically, we define a canonical field $x$ which can be related to the coordinates $z$ via the K{\"a}hler metric
\begin{equation}
d x^2 = K_{z \bar{z}} \left( dz_I^2 + dz_R^2 \right) .
\end{equation}
The distance between two points in field space is then found by integrating
\begin{equation}\label{eq:deltax}
\Delta x  = \int_{z^i}^{z^f} dz_I K_{z \bar{z}}^{1/2} \sqrt{ 1+ \left(\frac{dz_R}{dz_I}\right)^2}.
\end{equation}

The K{\"a}hler metric is approximately independent of the phase of $z$, and except in the near-conifold or near-LCS limits, falls off like $K_{z \bar{z}} \sim |z|^{-2}$ (see Sec.~\ref{sec:gentraj}). The path length for a curve around the conifold point at approximate distance $R$ can can therefore be approximated by $\Delta x \simeq R \sqrt{K_{z \bar{z}}} \Delta \Phi \simeq 0.1 R \Delta \Phi$. We can roughly determine the typical distance in field space between minima on adjacent sheets using $R=0.1$ and $\Delta \Phi \simeq 2 \pi$, yielding $\Delta x \sim 0.1$. This generic path length is only modified very close to the conifold and the large complex structure points, where the moduli space is curved. Thus, shorter paths can be obtained closer to the conifold, although in the near-conifold region $K_{z \bar{z}}$ diverges like $\ln R$.

In the flux landscape, we have $M^2 = M_p^2 / 2$, and we therefore expect the distance between adjacent minima to be of order $.1 M_p$. This implies (through the relatively large value of $\epsilon$) that gravitational effects will be important when calculating the instanton. With this in mind, we expect from the arguments in Sec.~\ref{sec:multscalars} that trajectories pass through the near-neighborhood of a HM point. Among the examples we have studied, the difference in energy between the false-vacuum minimum and the HM point is $V_N^{HM} - V_N^{F} \sim \mathcal{O}(0.1)$ (see Table~\ref{tab:examples}). The change in potential obtained by the monodromy transformation will be determined by the flux change $\Delta F_1$ and the period of the shrinking cycle $\Pi_1$. It is straight-forward to show that moderate changes in flux yield changes in energy of $\mathcal{O}(0.1)$ (again, see Table~\ref{tab:examples} for concrete examples). Since the barrier height is typically rather lo
 w compared to the vacuum energy splitting, the bubbles are in general not thin wall. Based on this information, we expect that the instanton action between de Sitter minima in our example flux-landscape will be of order 
\begin{equation}
S_I \sim M^4 / \mu^4 \propto \rho_I^3.
\end{equation}

\begin{figure*}[tb]
 \begin{center}
  \includegraphics[width=6.3cm]{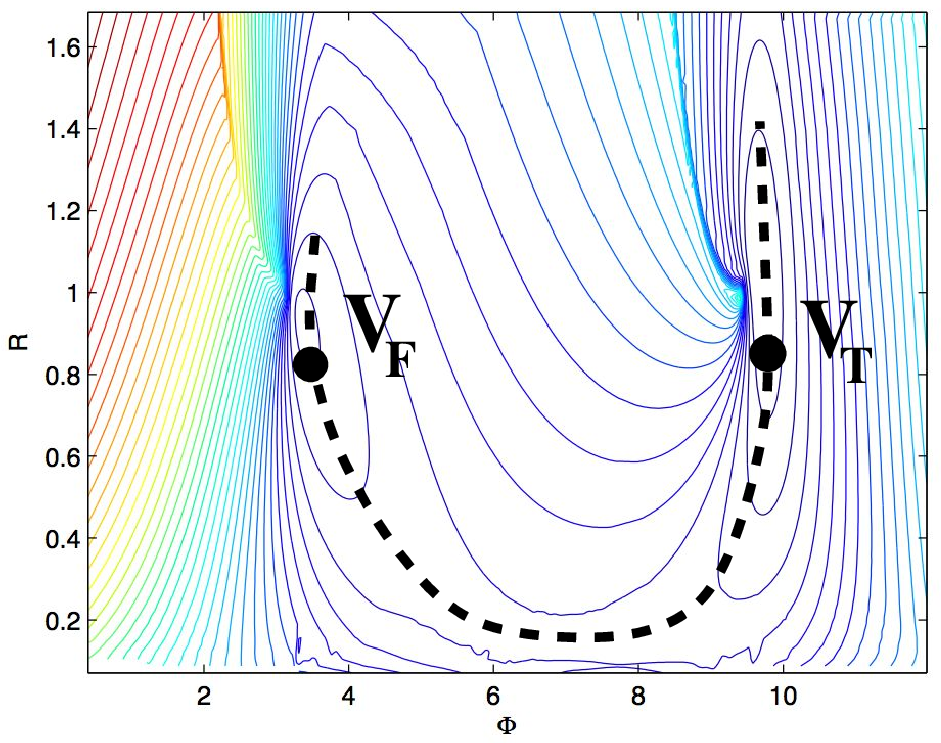}
  \includegraphics[width=11cm]{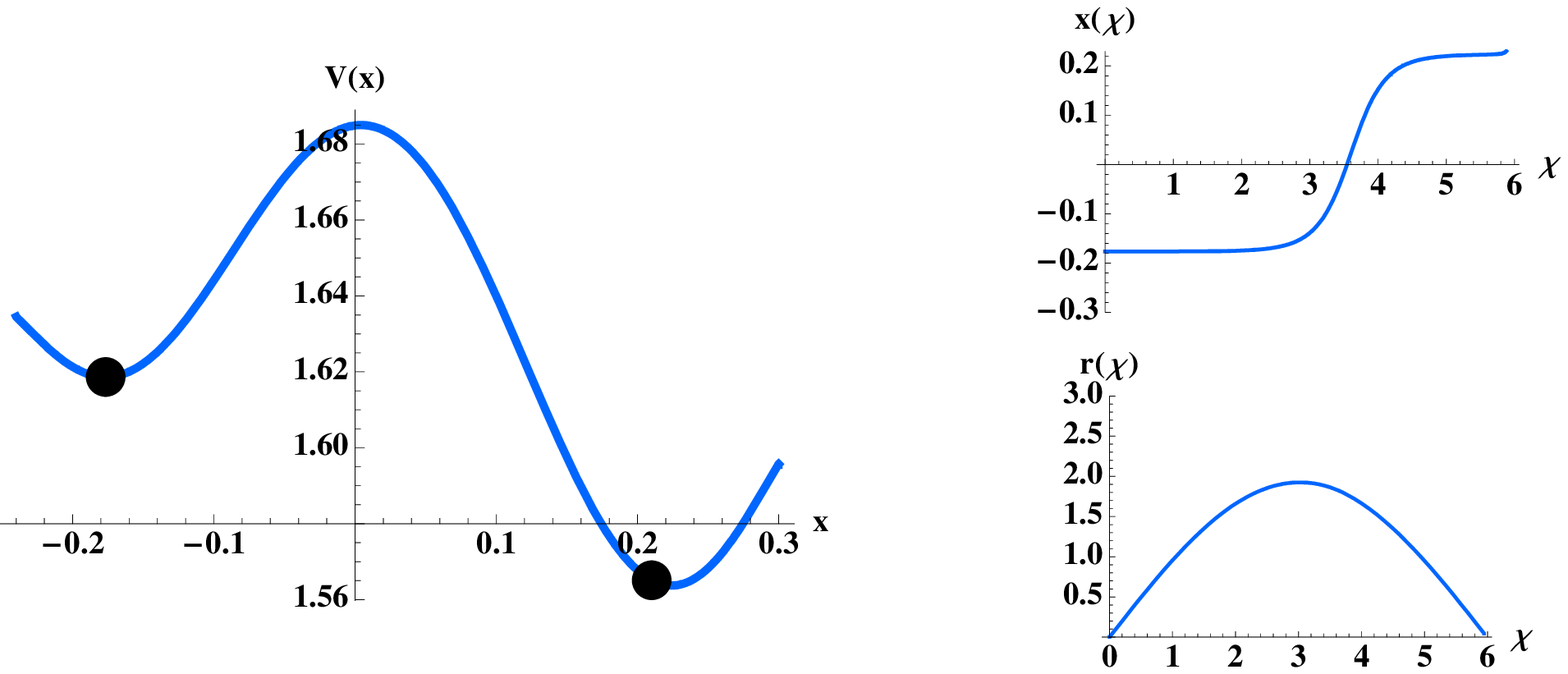} \\
  \includegraphics[width=6cm]{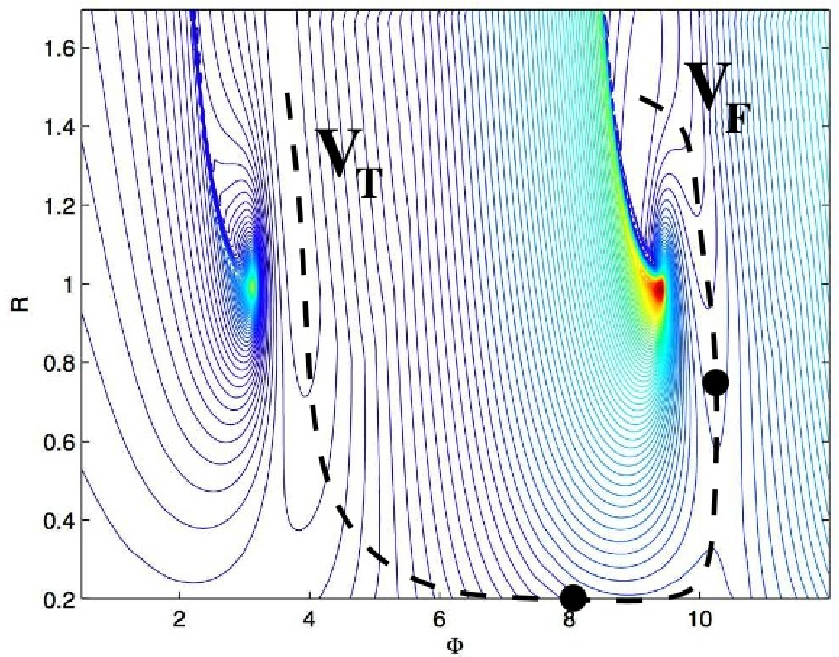}
\includegraphics[width=11cm]{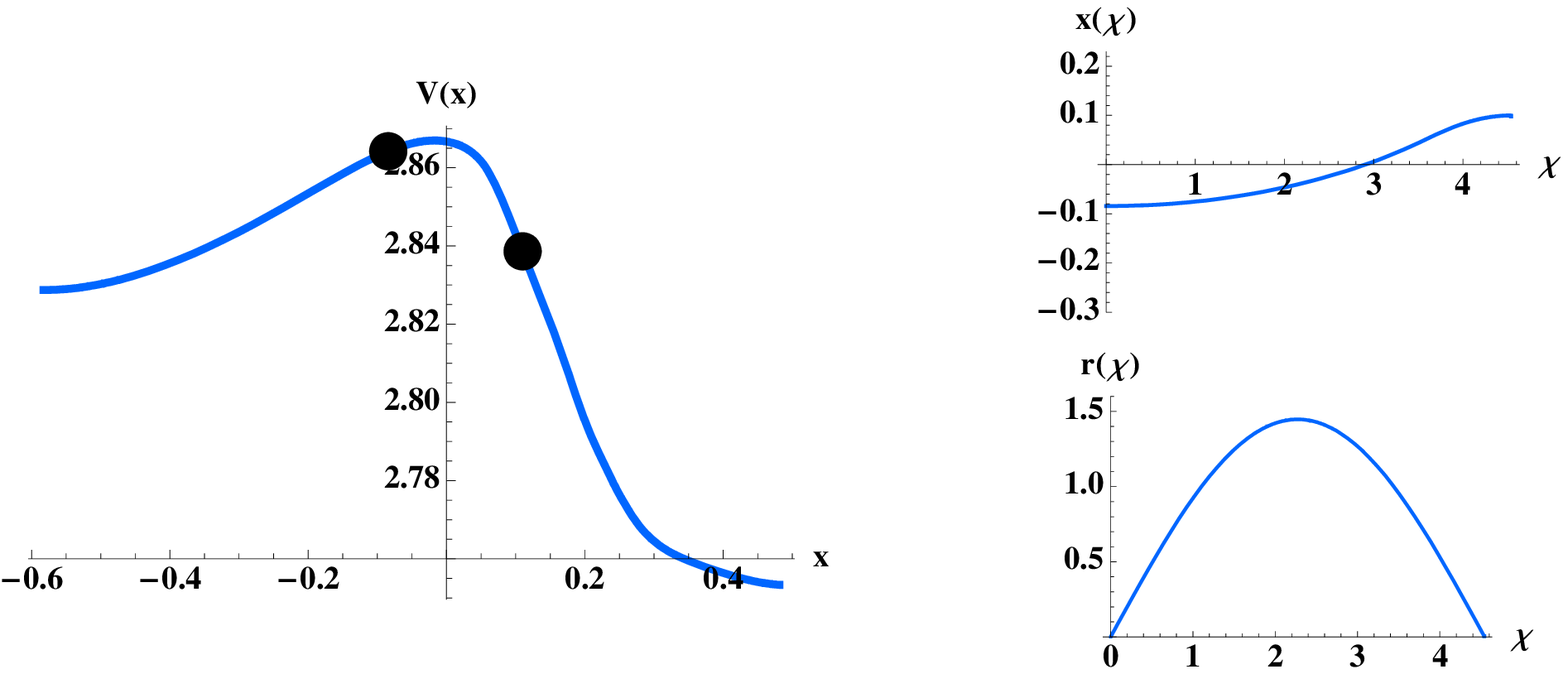}
 \end{center}
 \caption{Two minima with flux configurations $F=[0, -3, 5, 6];H=[1, 1, 1, 0]$ and $F=[-6, -3, 5, 6];H=[1, 1, 1, 0]$ (top) and $F=[3, -4, -1, -2]; H=[1, 0, 5, 0]$ and $F=[1, -4, -1, -2]; H=[1, 0, 5, 0]$ (bottom) connected by an instanton. On the far left, a contour plot of the potential in the polar representation is shown. The next cell depicts the potential obtained by sampling along the dashed line in the contour plot as a function of the canonical field $x$. The endpoints of the instanton connecting the two minima are denoted by the dots. On the far right, the instanton solution $x(\chi)$, $r(\chi)$ is displayed.}
 \label{fig:instanton1}
\end{figure*}

We can confirm this argument by studying the examples shown in Fig.~\ref{fig:instanton1}. Examining the features of each potential, we define a path between the two minima along which it is plausible that the instanton trajectory lies. We require that the path goes through the HM point and each minimum, and that it avoids any obvious run-away directions. Sampling the potential and K{\"a}hler metric along this path, and defining a canonical field through Eq.~\ref{eq:deltax}, we are left with a one-dimensional potential. The instanton can now be determined numerically as described in Sec.~\ref{sec:onescalar}. The chosen paths are depicted by the dashed lines in the contour plots of Fig.~\ref{fig:instanton1}, with the instanton endpoints denoted by the dots, and the explicit solutions $x(\chi)$ and $r(\chi)$ displayed on the far right. Small variations in the chosen path do not significantly change our results.

The instanton action can be computed from Eq.~\ref{eq:S_I}, yielding the results shown in Table~\ref{tab:actions}. Here, we also display the action for the HMinstanton. In both cases, we see that the action is $\sim {\mathcal O}(\rho_I^3)$ and the CDL instanton has lower action than the HM instanton, but not much lower, confirming our earlier arguments. 
 
\begin{table*}
\begin{ruledtabular}
\begin{tabular}{l c c c c c}
Upper Flux Configuration & Lower Flux Configuration & $\Delta F_1$ & $S_{I} - S_{BG}$ & $S_{HM} - S_{BG}$ \\
$F=[-6, -3, 5, 6];H=[1, 1, 1, 0]$ & $F=[0, -3, 5, 6];H=[1, 1, 1, 0]$ & 6 & $1.2 \rho_I^3$ & $13.4 \rho_I^3$ \\
$F=[3, -4, -1, -2];H=[1, 0, 5, 0]$ & $F=[1, -4, -1, -2]; H=[1, 0, 5, 0]$ & 2 & $3.1 \rho_I^3$ & $3.5 \rho_I^3$ \\
 \end{tabular}
 \end{ruledtabular}
 \begin{center}
 \caption{Instanton actions. \label{tab:actions}}
 \end{center}
\end{table*}

\subsection{Thin-wall instantons, BPS domain walls, and D5 branes}\label{sec:BPSD5}

In the previous section, we studied the transition between two de Sitter minima in the context of the no-scale flux-induced potential. Because the barrier height is fairly universal, and of order the energy splitting between the minima, these instantons are generically thick-wall. Including the K{\"a}hler sector, and stepping away from the no-scale potential, there will be Anti-de Sitter minima and de Sitter minima whose vacuum energy is not set by the fluxes. With the addition of these new scales, the vacuum energy splitting and the barrier height will no longer both be determined by the fluxes, and we can expect to find examples where the thin-wall approximation discussed in Sec.~\ref{sec:thinwall} applies. This will arise when considering near-BPS domain walls~\cite{Ceresole:2006iq}, as we now describe.
 
BPS domain walls separating two supersymmetric flux vacua (see~\cite{Cvetic:1996vr} for a review) are intimately related to thin-wall CDL bubbles, corresponding to the limit where the critical radius goes to infinity~\cite{Cvetic:1992st}. However, supersymmetric vacua are absolutely stable~\cite{Deser:1977hu,Witten:1981mf,Hull:1983ap}, and these domain walls cannot be interpreted as arising from tunneling transitions between vacua (physically, their infinite extent makes the production of such objects infinitely suppressed).

\begin{figure}[tb]
 \begin{center}
  \includegraphics[height=7cm]{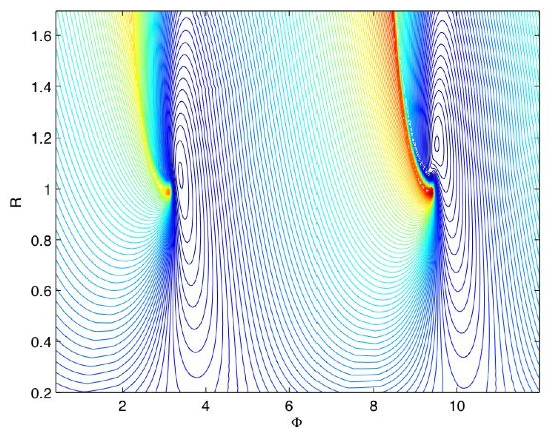}
 \end{center}
 \caption{Two zero-energy minima of the flux-induced potential for the Mirror Quintic. Although the minima on the two sheets are connected by a smooth potential there is no tunneling between them.}
 \label{fig:susy}
\end{figure}

In the no-scale potential, a necessary (though not sufficient, since we must also include the K{\"a}hler sector) condition for a supersymmetric vacuum is that the potential is zero at the minimum. This follows from the vanishing of the F-terms $D_z W = D_{\tau}W=0$. It is easy to find continuously connected vacua that fulfil $D_z W = D_{\tau}W=0$ on the Mirror Quintic  and an example of two such vacua is shown in Fig.~\ref{fig:susy}. Although these minima are connected, they are absolutely stable; since the vacuum energies are the same the tunneling rate is infinitely suppressed.

Following~\cite{Ceresole:2006iq}, if the potential energy of one of the minima were lifted by a small amount (using ingredients such as branes to do so in a controlled way), the minimum would no longer be totally stable. Provided that this uplift is small, the thin-wall analysis could be applied and the tension of the bubble computed using Eq.~\ref{eq:tension}. Sampling the potential in Fig.~\ref{fig:susy} along a variety of paths between the two minima and transforming to a canonically normalized field using Eq.~\ref{eq:deltax}, we obtain by numerical integration
\be
\sigma=\sigma_N M \mu^2  = 0.1M \mu^2 = 0.1 M_p^3 \rho_I^{-3/2}.
\label{eq:Ntension}
\ee
Note that the numerical factor $\sigma_N$ includes a factor of $\tau_I^{-1/2}$ evaluated along the path. The exact value of $\sigma_N$ changes by a few percent from path to path (and from potential to potential), but this result simply follows from the universality of the distance between the minima and the barrier heights that were discussed above. We can therefore expect that this will be the general scale of the tension of thin-wall bubbles in flux-changing transitions. This result agrees with the scaling presented in Ref.~\cite{Dine:2007er}. 

Once the vacuum energies of the true and false vacua are known, the bounce action can be determined from Eq.~\ref{eq:thinwallaction} using the tension Eq.~\ref{eq:Ntension}. The tension scale, as obtained from the scale of the potential, agrees with the scale of the BPS domain wall tension as well
\be
\sigma_{dw} = \Delta \left( e^{K/2} |\hat{W}| \right) \sim M_p^3 \rho_I^{-3/2} g_s^{1/2}.
\ee
Here $\hat{W}$ is the standard supergravity superpotential of mass dimension 3, in contrast to $W$ used in previous sections.

In this paper, we have used the fact that monodromy transitions yield continuous potential barriers between minima with different flux configurations.  Another approach to transitions between vacua with different flux is the nucleation of charged brane bubbles~\cite{Brown:1988kg}. In string theory, D- and NS-branes source flux and can act as infinitesimally thin domain walls between vacua with different flux \cite{Gukov:1999ya}. It has therefore been proposed that tunneling can proceed by the nucleation of D- or NS-brane bubbles \cite{Feng:2000if, Bousso:2000xa,Frey:2003dm}. If these branes wrap internal three-cycles of the manifold, they will appear as two-dimensional domain walls in the four-dimensional world. The tension of the effective bubbles is given by
\be
\sigma_{D5} = \mid  \Pi_i \mid M_p^3 \rho_I^{-3/2} g_s^{1/2},
\ee
where $\Pi_i$ is the period over the wrapped three-cycle. 

It is interesting to see that the wrapped brane tension agrees roughly with the thin-wall tension in Eq.~\ref{eq:Ntension}. With this in mind, it is natural to consider that the potential barrier between the two minima in Fig.~\ref{fig:susy} is a resolution of a D-brane domain wall that separates the minima. By continuity, it would be unnatural if the domain walls derived from the supergravity potential in the thin-wall limit had a radically different interpretation than the thick-wall configurations studied above. Therefore we are lead to conclude that the thick-wall instantons that interpolate between minima with badly broken supersymmetry also have a interpretation as resolved D-branes, as suggested in~\cite{Danielsson:2006xw}. However, the field configuration or resolved brane interpolating between these minima is smeared over nearly a Hubble volume of the false vacuum.

Interestingly, the smearing of the brane is unrelated to its charge~\footnote{Thus, the smeared D-branes do not correspond to the `fat brane' interpretation of stacks of D-branes suggested by~\cite{Lizzi:1997tv}.}. In the monodromy staircase, the change in flux between two potential sheets is set by the flux through the shrinking cycle. This need not be one. In the D-brane nucleation picture, each brane has charge one and it is necessary to nucleate several brane bubbles to change the flux configuration by several units. It would seem that the action for such repeated nucleations is large. On the contrary, the actions computed above are, for reasonable fluxes, practically independent of the change in flux. In fact, in the two examples presented in Fig.~\ref{fig:instanton1} it is clearly seen that the transition with $\Delta F = 6$ is faster than the transition with $\Delta F = 1$. It has been argued that an enhancement of the nucleation rate is achieved when including gauge f
 ields associated with the brane stack in the pre-factor~\cite{Feng:2000if} (see also~\cite{Garriga:2003gv}). It is unclear how to make a connection with this picture in the present scenario.  

Finally, although the brane nucleation picture is natural from a stringy perspective, it has been questioned. Static branes can be described in the supergravity approximation of string theory~\cite{Cvetic:1996vr}, but it is difficult to formulate a consistent action to describe brane nucleation~\cite{deAlwis:2006cb}. Of course, these complications do not forbid brane nucleations in the landscape, but make it difficult to describe such transitions (and calculate their probability) in the low energy supergravity approximation. Here we see that there is a complementary approach to transitions between flux minima, that is valid as long as the low-energy effective field theory describing moduli stabilization is. We leave a more detailed analysis of the similarities and differences between these two points of view to future work.

\section{Eternal inflation and the landscape}
\label{sec:et_inf}

Thus far, we have used the monodromies of the complex structure moduli space to explicitly construct the scalar potential in the vicinity of and between different flux vacua. In the limit of large compactification volume (which we assume) and weak string coupling (which we explicitly verify), we have found that the decay probability per unit four-volume is exponentially suppressed. Since this decay probability will be much smaller than $H_F^{4}$ (the Hubble constant in a false vacuum), then a vacuum, once accessed, will exhibit a phenomenon known as eternal inflation (see e.g.~\cite{Aguirre:2007gy,Guth:2007ng,Winitzki:2006rn} for a review). The large-scale structure of the universe then consists of many causally disconnected regions in which different vacua might be physically realized, but where there is always a spatial slicing along which part of the universe remains in the false vacuum. Our present vacuum presumably exists in some local region of the eternally inflating u
 niverse, and it is plausible (though not logically necessary), and possibly overwhelmingly likely~\footnote{This assumes some measure over cosmological histories; the measure problem for eternal inflation is a controversial topic with no clear resolution to date.}, that our universe was formed in the aftermath of a transition between two vacua. 

Such transitions can result in local regions of the universe undergoing a period of slow roll inflation, reheating, and subsequently (in a vacuum yielding the standard model of particle physics) produce the Standard Big Bang (SBB) cosmology. This model goes by the name of open inflation~\cite{Gott:1982zf,Bucher:1994gb}, and the "big bang" is replaced by an infinite null surface on which the initial conditions for the inflationary epoch are set by the endpoints of the CDL instanton mediating the transition. In principle, this allows us to understand both the inflationary epoch immediately preceding SBB cosmology {\em and its initial conditions} entirely in terms of our four-dimensional effective theory~\footnote{This assumes that eternal inflation can be adequately described by the semi-classical theory of scalars coupled to gravity. This semi-classical description may be inadequate and misleading~\cite{ArkaniHamed:2007ky}}. 

In our example landscape, we have neglected a number of elements that will play important roles in a realistic picture of the resulting eternally inflating spacetime. Most importantly, we have not explicitly stabilized the K{\"a}hler moduli, assuming that this can be accomplished at large compactification volume by considering corrections to the K\"ahler potential and superpotential. Because most of the flux vacua we are considering break supersymmetry at tree level, this may be difficult (or impossible) to achieve in general~\cite{Dimofte:2008jg}. We will have no more to say about this issue, but note that in order to connect the string theory landscape to false-vacuum eternal inflation, one must verify the existence of consistent de Sitter vacua at a relatively high energy scale.

The scalar potential for the K{\"a}hler moduli always goes to zero at infinite volume, and so any minima we might find will be unstable to spontaneous decompactification~\cite{Giddings:2004vr} to ten non-compact dimensions. This is mediated by the same CDL instantons discussed in Sec.~\ref{sec:tunneling} as long as the instanton endpoints are at small enough volume for the effective four dimensional theory to be valid. As decompactification occurs, the scale of the flux-induced potential will decrease and the massive KK modes we have neglected will become light, inducing significant corrections, and invalidating our effective theory. The amplitude for such transitions, while typically gravitationally suppressed~\cite{Westphal:2007xd}, is somewhat model dependent, precluding a direct comparison with the rates we have found for flux-changing transitions. 

There will be other orthogonal directions of instability as well, perhaps connecting vacua through the K{\"a}hler~\cite{Escoda:2003fa} or open string sectors~\cite{Gomis:2005wc}. In more general compactifications, we will also have other complex structure moduli, adding the potential of many more directions of instability to flux changing transitions than the two (associated with the complex field $z$) we have considered here. Typical vacua will therefore have many decay channels, in which the local flux configuration, size of various cycles or position of branes in the internal manifold, or even the overall scale of the compactification change. Because of the potentially large number of ways to decay, one might worry that the total decay probability per unit four-volume out of a given vacuum becomes greater than $H_F^{4}$, and the phase transition completes, ending eternal inflation. However, this would require atypically high transition rates or an exponentially large numbe
 r of decay channels, neither of which are obviously present.

Once one moves away from the no-scale potential that we have studied in our example landscape, there will be many Anti-de Sitter (AdS) vacua as well as zero and positive energy vacua. Transitions into the basin of attraction of an AdS vacuum results in a big crunch, and near the singularity our effective theory will break down. These vacua act as sinks~\cite{Ceresole:2006iq} for probability current, and cause the fraction of comoving volume in de Sitter vacua to monotonically decrease. The decompactification transitions discussed above would seem to act similarly, removing four-dimensional co-moving volume from the realm where the effective theory applies, although it is unclear how one is to define a sensible measure for an eternally inflating spacetime in which the dimensionality of space effectively changes. Given that such transitions will certainly occur, and that the existence of sinks for the probability current drastically affect the resulting distributions (see, eg~\
 cite{Linde:2006nw}), it seems crucial to include them at least in a primitive manner if existing measure proposals are to apply to a description of the string theory landscape.

In sum, we expect that if eternal inflation occurs, then pockets of the eternally inflating spacetime can be described by the four-dimensional low energy effective theory of the full scalar moduli potential coupled to gravity. Eternal inflation is not guaranteed to be past eternal~\cite{Borde:2001nh} (see however~\cite{Aguirre:2003ck}), and so some unknown process will set the ultimate initial conditions. The effective theory can then be used to describe transitions between low-energy vacua, but it is possible for it to dynamically bring about its own demise, through the formation of singularities and decompactification transitions. This patchwork makes it difficult to assess the ultimate validity of the theory, but the possibility that the low-energy theory can encapsulate the transition into our present vacuum and our observed cosmology is intriguing. 

\section{Conclusions}
\label{sec:conclusions}

The use of monodromies to generate the potential between flux vacua has allowed us to explore both  cosmological field dynamics and tunneling in a landscape of Type IIB flux compactifications. Couching the description of transitions between flux vacua in the language of a four-dimensional effective theory of scalar fields coupled to gravity yields an explicit picture of the way in which vacua of the string theory landscape are connected, and our example of the Mirror Quintic Calabi--Yau has yielded some quantitative information on the features of such a landscape. Because this analysis has been rather lengthy, we summarize our main results below

\begin{itemize}
\item Monodromies can be used to find a continuous potential between two different flux vacua. Not all flux configurations can be reached in this way, and adjacent minima need not differ by one unit of flux.
\item We have considered the dynamics of the complex structure modulus and axio-dilaton for the Mirror Quintic Calabi--Yau, finding that the slow-roll conditions are typically not satisfied by the flux-induced potential, except for highly non-generic situations where an approximate inflection point could arise due to the merging of a minimum and a saddle point. For homogenous and isotropic cosmological solutions,  the field trajectories do not exhibit chaotic behavior and separate neatly into basins of attraction of the various minima. 
\item The barrier height and energy splitting between adjacent de Sitter minima are fairly universal and of order $\Delta V_N \sim .1$, implying that the instanton mediating the transition will be thick-wall. In addition, the canonical field distance between adjacent vacua is fairly universal and of order $M_p$, implying that gravitational effects will be important. We therefore estimate the transition rate between de Sitter vacua with moderate flux integers as $\Gamma \sim \exp ( - g_s^{-1} \rho_I^{3})$, which is confirmed by the construction of explicit examples. For large volume compactifications and weak string coupling, the rate is exponentially suppressed.
\item We can imagine that there will be thin-wall instantons when supersymmetry is unbroken in the flux sector, in which case the universal properties of the barriers between zero-energy vacua allow us to estimate the tension of the bubble wall for flux-changing transitions to be very close to $\sigma \simeq .1 M_p^3 \rho_I^{-3/2}$.
\item The tension of these domain walls scales the same as the tension of a D- or NS-brane bubble, suggesting that domain walls constructed using the scalar potential we have computed are resolved branes or stacks of branes, as was first suggested in~\cite{Danielsson:2006xw}. By continuity, we expect this interpretation to carry over to the description of transitions between de Sitter minima. The bubble wall is in this case highly non-localized, suggesting a smearing of D-branes in de Sitter space. Since monodromy transformations can connect configurations differing by more than one unit of flux, the unit change implied by the brane-nucleation picture is rather misleading.
\end{itemize}

Let us now discuss some limitations and extensions of our results. The transitions between vacua related by monodromies are expected to remain in string theory models with all moduli fixed. In particular, this should be true in Type IIB compactification on conformal Calabi--Yau manifolds, whose internal geometry is largely determined by periods, that are affected by monodromies. Most Calabi--Yau manifolds will have more complex structure moduli, and more monodromies, thus connecting larger sets of vacua. As a first approximation, the Mirror Quintic results can be used to estimate the tunneling probability between vacua in these models. 

We have seen that the tunneling probability between supersymmetry-breaking flux vacua is exponentially suppressed. This suppression is minimized by diminishing the internal volume and increasing the string coupling. However, these regions of the landscape are not well described by the effective field theory we are using (this complements the realm of validity of the semi-classical approximation to the bounce action, which requires $M^4 \gg \mu^4$), and so we should tread carefully in this regime. Alternatively, a very large flux would yield a fast transition between minima (due to an overall increase in $V_N$ over the instanton trajectory). This is also problematic, since such large flux leads to a radical back reaction on the internal geometry, and eventually the internal manifold is no longer conformally Calabi--Yau (see~\cite{Grana:2005jc} and references therein). It is unclear if the necessary monodromy transformations will survive in this limit. In addition, the overall 
 scale of the potential will become much higher than the string and Kaluza-Klein scales, seriously compromising the effective theory.

We have also concluded that the generic distance between minima in the monodromy staircase is large ($\mathcal{O}(M_p)$). This is problematic for chain inflation~\cite{Freese:2004vs,Freese:2006fk}, which is only allowed if the inter-minimum distance is much shorter~\cite{Chialva:2008zw}. Nevertheless, there might be tuned scenarios where the minima and the intermediate saddle point are brought closer together. For example, the extrema could lie closer to the conifold point than the examples studied here, opening up for shorter paths.

Furthermore, we have argued that thin wall instantons could exist in monodromy staircases when zero-energy minima are uplifted by effects in the open string or K{\"a}hler sectors. The relative universality of the bubble-wall tension allows a determination of the tunneling rates from Eq.~\ref{eq:thinwallaction} once the vacuum energies are known. This can be compared with the analysis of~\cite{Ceresole:2006iq,Dine:2007er}, where such thin-wall transitions were considered. In reality, there will most likely be a spectrum between thin- and thick-wall solutions, determined by the hierarchy of scales between the barrier height and the vacuum energy splitting. 

In order to support eternal inflation (or at least the variety that is relevant to our cosmological history, including an epoch of open inflation), transitions in the landscape from relatively high energy de Sitter minima must be considered. On the other hand, for assessing the stability of our present vacuum, one is typically interested in transitions from a low-energy de Sitter vacuum to a big crunch (towards an AdS minimum). It is important to establish the existence and properties of both types of transitions. Transitions among the supersymmetry-breaking flux vacua of Sec.~\ref{sec:Eucpaths} offer an example relevant to the first scenario, and the near-BPS bubbles of Sec.~\ref{sec:BPSD5} will be relevant for the second. However, once the K{\"a}hler moduli are re-introduced, more work must be done to  establish the continued existence of the supersymmetry-breaking flux vacua, and the particulars of uplifting the zero-energy vacua.

It can be hoped that with a better understanding of transitions in the landscape, we will be able to determine the connection of the string theory landscape to eternal inflation, and ultimately to our observed cosmology through the phenomenology of open inflation. There are many open questions in this direction, and we leave further discussion to future work.

\section*{Acknowledgments}
This work was initiated during the 2007 Les Houches school "String Theory and the Real World", and we would like to thank the organizers, lecturers and fellow students of that school. We would also like to thank T. Banks, F. Denef, M. Dine, G. Festuccia, N. Johansson, A. Morisse, B. Underwood and K. Vyas for interesting discussions and U. Danielsson and T. Dimofte for discussions and comments on the manuscript. ML would like to thank Caltech and MJ would like to thank Uppsala University for their generous hospitality during important stages of this work. MJ acknowledges support from the Gordon and Betty Moore Foundation.

\appendix
\section{Notation, geometry and scales}
\label{ApB}

\subsection{Geometry}
Calabi--Yau manifolds are complex, Ricci-flat manifolds. The complex structure moduli (CSM) space associated to a Calabi--Yau manifold parametrize the different ways one may split the real coordinates on the manifold into holomorphic and anti-holomorphic coordinates. On every Calabi--Yau manifold there is a non-vanishing holomorphic form $\Omega$ of middle cohomology that encodes this choice. E.g. on a six-dimensional Calabi--Yau there is a holomorphic three-form $\Omega \in H^{(3,0)}$ that is key to understanding the features of its CSM space.

In particular, by choosing a basis of $N$ three-cycles $C_I$ for the homology class $H_3$ we can define the periods $\Pi_I$ over these cycles as
\be
\Pi_I = \oint_{C_I} \Omega.
\ee
These periods measure the `holomorphic volumes' of the three-cycles and determine the geometry of the CSM space through the K{\"a}hler potential
\begin{eqnarray}\label{eq:kahler}
K_{\mathrm{cs}}(z) &=& -\ln \left( -i\int_{CY} \Omega \wedge \bar{\Omega} \right) \nonumber \\  &=& -\ln \left( -i\Pi^{\dagger}(z) \cdot Q^{-1} \cdot \Pi(z) \right).
\end{eqnarray}
For notational simplicity, we suppress the  $\bar{z}$ dependence of $K_{\mathrm{cs}}(z)$. Here we have used the period vector
\begin{equation}
\Pi(z)=\left(
\begin{array}
[c]{c}%
\Pi_{1}(z)\\
\Pi_{2}(z)\\
\vdots\\
\Pi_{N}(z)
\end{array}
\right) ,
\end{equation}
where $z$ is an $(N/2-1)$-dimensional (complex) coordinate on the CSM space. $Q_{IJ}$ is the intersection matrix of the three-cycles given by
\begin{equation}
Q_{IJ} =  C_I \cap C_J.
\label{eq.Q}
\end{equation}
The three-cycle basis $C_I$ is called canonical if the cycles intersect pairwise with intersection number $\pm 1$. These geometric features of the CSM space have been derived using special geometry, see \cite{Candelas:1990pi,Strominger:1990pd,Dixon:1989fj}, which is applicable for $\cN=2$ vector multiplets. The structure survives in $\cN=1$ orientifolds of these models, such as the ones studied here. The expression for the K{\"a}hler potential is valid if the Calabi--Yau volume is big, which we assume in this paper. If the volume is small, the K{\"a}hler potential is modified by warping~\cite{DeWolfe:2002nn,Giddings:2005ff,Shiu:2008ry,Douglas:2008jx}.

\subsection{Flux compactification and the scalar potential}
We now turn to flux compactifications. Compactifying Type IIB string theory on a conformally Calabi--Yau manifold gives an effective four-dimensional theory. Turning on three-fluxes (see below) breaks supersymmetry to $\cN = 1$ and generates a scalar potential that depends on the complex structure moduli $z$, the K{\"a}hler volume modulus $\rho = \rho_R + i \rho_I$ and the Type IIB axio-dilaton $\tau = \tau_R + i\tau_I$:
\be
V = e^K \left( K^{a\bar{b}} D_aWD_{\bar{b}} \overline{W}  - 3 W ^2 \right),
\ee 
where the indices go over $z, \rho$ and  $\tau$. The scalar potential is determined in terms of a Gukov--Vafa--Witten superpotential $W$ and a K{\"a}hler potential $K$. The equation also includes the covariant derivatives $D_i = \partial_i +  \partial_iK$ and the K{\"a}hler metric $K_{i\bar{j}}=\partial_i\partial_{\bar{j}}K$. $K$ is given by
\be
K=-\ln\left(  \tau_I \right)  +K_{\mathrm{cs}}\left(  z,\bar
{z}\right)  -3\ln\left(  \rho_I \right),
\ee
where $K_{\mathrm{cs}}$ was defined above. The superpotential depends on the three-form fluxes, the axio-dilaton and the periods.
\begin{eqnarray}\label{eq:superpotential}
W &=& \frac{1}{(2 \pi)^2\alpha'} \int_{CY}\Omega\wedge(F_{(3)}-\tau H_{(3)}) \nonumber \\
&=& (F-\tau H)\cdot \Pi.
\end{eqnarray}
Here the three-form fluxes have been collected in row vectors $F=[F_1,F_2,F_3,F_4]$ and similarily for $H$. The Dirac quantization conditions for the fluxes then become
\begin{eqnarray}
\int_{C_J} F_{(3)} &=& -(2 \pi)^2\alpha'  F_I Q_{IJ} \mbox{ , } \nonumber \\
\int_{C_J} H_{(3)} &=& -(2 \pi)^2\alpha'  H_I Q_{IJ}.
\label{eq:quant}
\end{eqnarray}
This unconventional definition of the flux vectors is used get a simpler expression for the superpotential.

Since $W$ is independent of $\rho$, it is straight-forward to show that the potential is of no-scale type, i.e. that
\be \label{eq:dimlessv}
V = e^K \left( K^{z\bar{z}} D_zWD_{\bar{z}} \overline{W}  + K^{\tau\bar{\tau}} D_{\tau}WD_{\bar{\tau}} \overline{W} \right).
\ee 
The potential is positive semi-definite, and as presently defined is a dimensionless function (we discuss the overall scale of the potential in the next subsection). Minima that preserve supersymmetry in the flux sector have vanishing F-terms $D_zW= D_{\tau}W=0$ and zero potential energy.

\subsection{Dimensional reduction and energy scales}
To discuss physical properties, the correct scale of the potential is needed. To restore the scaling factors, we will briefly review the dimensional reduction from ten to four dimensions. We will follow \cite{deAlwis:2006am, Giddings:2001yu, DeWolfe:2002nn}, where this reduction has been thoroughly discussed.

The bosonic low-energy action for Type IIB string theory in the ten-dimensional Einstein frame is given by (up to subleading $\alpha'$ and $g_s$ corrections)
\begin{widetext}
\be
S_{IIB} = \frac{1}{(2\pi)^7 \alpha'^4} \int d^{10}x \sqrt{-\cG} \left( \cR -\frac{\partial_M\tau \partial^M\bar{\tau}}{2 \tau_I^2} - \frac{ G_{(3)}\cdot \bar{G}_{(3)} }{12 \tau_I} - \frac{|\tilde{F}_5|^2}{4\cdot 5!} \right) 
- \frac{1}{4 i (2\pi)^7 \alpha'^4} \int \frac{C_{(4)} \wedge G_{(3)} \wedge \bar{G}_{(3)}}{\tau_I}. 
\label{eq:10Daction}
\ee
Here $G_{(3)}=F_{(3)}-\tau H_{(3)}$ is the three-form flux, $\tilde{F}_5$ is a self-dual five-form flux, $\tau=i g_{s}^{-1}+C_{0}$ is the axio-dilaton. We compactify on a (conformal) Calabi--Yau threefold, and use the ten-dimensional block-diagonal metric ansatz:
\be
ds_{10}^2 = \cG_{MN}dz^Mdz^N = e^{-6u(x)}e^{2A(y)}g_{\mu\nu}dx^{\mu}dx^{\nu} + e^{2u(x)}e^{-2A(y)}\tilde{g}_{mn}dy^{m}dy^{n},
\label{eq:metric}
\ee
\end{widetext}
where an overall scale factor of the internal volume, $e^{2u}=  \rho_I^{1/2}$, and a warp factor, $e^{-2A}$, has been factored out from the six-dimensional metric, leaving $\tilde{g}_{mn}$ which is Calabi--Yau. The four-dimensional part of the metric is warped, with warp factor $e^{2A(y)}$. In the following, we will assume that this warp factor can be neglected. For a discussion of the effects of the warp factor on the low energy effective field theory, see \cite{Giddings:2001yu, DeWolfe:2002nn,Giddings:2005ff,Shiu:2008ry}. Choosing this metric implies that our computations are made in the four-dimensional Einstein frame, as is seen below.

The terms in the ten-dimensional action will contribute to various terms in the four-dimensional action upon dimensional reduction. The Ricci scalar yields the four-dimensional Einstein--Hilbert term, kinetic terms for the geometric moduli and a contribution to the no--scale potential. The flux terms will also contribute to the potential. The kinetic term for the axio-dilaton in the ten-dimensional action gives a similar kinetic term in the four-dimensional effective field theory. As shown in \cite{Giddings:2001yu, DeWolfe:2002nn}, we get the four-dimensional Einstein--Hilbert term
\be
\begin{split}
&\frac{1}{(2\pi)^7 \alpha'^4}  \int d^{10}x \sqrt{-\cG} \cR \\
&= \frac{1}{(2\pi)^7 \alpha'^4}  \int d^{4}x \sqrt{-g_4} R^{(4)} \int d^{6}y \sqrt{\tilde{g}_6}+ ...\\
&= \frac{M_p^2}{2} \int d^{4}x \sqrt{-g_4} R^{(4)} + ...
\end{split}
\ee
The last equality defines the four-dimensional Planck scale
\be
\frac{M_p^2}{2} =\frac{\tilde{\cV}}{(2\pi)^7 \alpha'^4} = \frac{1}{2 \pi \alpha'},
\ee
where $\tilde{\cV} = \int d^{6}y \sqrt{\tilde{g}_6}$ is the volume of the internal manifold when $\rho_I = 1$. Since all volumes are measured in units of $\alpha'$,  we can conveniently choose  $\tilde{\cV}=(2 \pi)^6 \alpha'^3$ \cite{deAlwis:2006am}, yielding the second equality. Note that this four-dimensional Planck mass is measured in the four-dimensional Einstein frame, where the effective string  and Kaluza-Klain scales are space-time dependent. Neglecting warping, the string scale is given by  $M_s^2= ( 2\pi\alpha' \tau_I^{1/2} \rho_I^{3/2})^{-1}$ and the lightest Kaluza--Klein mode is  at $M_{KK}^2=(2\pi\alpha' \rho_I^2)^{-1}$~\cite{deAlwis:2006am}. 

The potential term arising from $\cR$ and the fluxes is given by~\cite{Giddings:2001yu, DeWolfe:2002nn}
\be
S_V = \frac{1}{(2\pi)^7 \alpha'^4} \int d^4 x \sqrt{-g_4} \int \frac{1}{\tau_I \rho_I^3} G_{(3)}^+ \wedge \star_6 \bar{G}_{(3)}^+,
\ee
where $G_{(3)}^{\pm} = \frac{1}{2} (G_{(3)} \pm i \star_6 \bar{G}_{(3)})$ and $\star_6$ is the six-dimensional Hodge star operator. The 10-dimensional equations of motion imply that $G_{(3)}$ is harmonic on the Calabi--Yau \cite{Giddings:2001yu}. Thus, we can expand it in a basis of harmonic 3-forms
\begin{eqnarray} 
G_{(3)} &=& \left(\int \Omega \wedge \bar{\Omega} \right)^{-1} \\ &\times& \left[ \Omega \int G_{(3)} \wedge \bar{\Omega} + G^{\alpha \bar{\beta}} \bar{\chi}_{\bar{\beta}} \int G_{(3)} \wedge \chi_{\alpha} \right] \nonumber.
\end{eqnarray}
Here $\Omega$ is the harmonic (3,0)-form, $\chi_{\alpha}$ are $h^{(2,1)}$ harmonic (2,1)-forms and $G^{\alpha \bar{\beta}}$ is the inverse of the metric $G_{\alpha \bar{\beta}} = \left(\int \Omega \wedge \bar{\Omega} \right)^{-1} \int \chi_{\alpha} \wedge \bar{\chi}_{\bar{\beta}}$. Assuming that $\rho$ and $\tau$ is independent of the internal coordinates, the potential term can be written
\begin{widetext}
\be
S_V = \frac{1}{(2\pi)^7 \alpha'^4} \int d^4 x \sqrt{-g_4} \frac{1}{ \rho_I^3 \tau_I \int \Omega \wedge \bar{\Omega}} \left[\int G_{(3)} \wedge \bar{\Omega} \int \bar{G}_{(3)} \wedge \Omega + G^{\alpha \bar{\delta}} \int G_{(3)} \wedge \chi_{\alpha} \int \bar{G}_{(3)} \wedge \bar{\chi}_{\bar{\delta}}. 
\right]
\label{eq:pot}
\ee
\end{widetext}

To rewrite this in $\cN =1$ form, we first identify the K\"ahler potential defined above: $e^K = ( \rho_I^3 \tau_I \int \Omega \wedge \bar{\Omega} )^{-1}$. Furthermore, since the fluxes are quantized with respect to the three--cycles they thread (see Eq.~\ref{eq:quant}), the terms in the brackets in Eq.~\ref{eq:pot} scale as $(2 \pi)^4 \alpha'^2$. Using the dimensionless Gukov--Vafa--Witten superpotential defined in Eq.~\ref{eq:superpotential} the potential is given by
\begin{eqnarray} \label{eq:V_scale}
V &=& \frac{ (2 \pi)^4 \alpha'^2 } {(2\pi)^7 \alpha'^4} e^{-K} \left( K^{i\bar{j}}D_iWD_{\bar{j}}\overline{W}\right)  \nonumber \\ 
&=& \frac{M_p^4}{\pi} \frac{g_s}{\rho_I^3} v (z, \tau),
\end{eqnarray}
where the index $i,j$ runs over the complex structure modulus $z$ and the axio-dilaton $\tau$ and 
\begin{equation}
v (z, \tau) =   e^{K_{cs}} \left( K^{i\bar{j}}D_iWD_{\bar{j}}\overline{W}\right).
\end{equation}

The dimensional reduction also yield the kinetic terms for the geometric moduli and the axio-dilaton \cite{Giddings:2001yu, DeWolfe:2002nn}:
\be
M_p^2\int d^4x \sqrt{-g} \left(  -  K_{i \bar{j}} \partial_{\mu}\psi^i \partial^{\mu}\bar{\psi}^{\bar{j}}  \right),
\ee
where $K_{i\bar{j}} = \partial_{\psi_i}\partial_{\bar{\psi}_{\bar{j}}}K$ is the (positive definite, block diagonal) K{\"a}hler metric on $\psi$ space computed above. The fields $\psi_i$ are the axio-dilaton and the complex structure and K{\"a}hler moduli. 

Alternatively, to confirm with standard  $\cN =1$ notation, we can define  dimensionful K\"ahler and superpotentials as 
\be
\hat{K} = K M_p^2 \mbox{ , } \hat{W} = W M_p^3.
\ee
This redefinition still yields a potential of mass dimension 4, but redefines the kinetic term to a canonical form. In this paper, we will mainly use the dimensionless quantities, which can be computed numerically.

\bibliography{landscapecosmo}
\end{document}